\documentclass[epj,nopacs]{svjour}

\usepackage{amssymb}
\usepackage{epsfig}
\usepackage{eucal}
\usepackage{amsmath}

\newcommand{\beq}{\begin{equation}} 
\newcommand{\eeq}{\end{equation}}   
\newcommand{\bea}{\begin{eqnarray}} 
\newcommand{\eea}{\end{eqnarray}}
\newcommand{\non}{\nonumber}
         
\newcommand{\Eq}[1]{Eq.~(\ref{#1})}

\def\Li{\hbox{Li}}  
                                  
\headnote{
\hfill \parbox{4cm}{\tt \normalsize CERN-PH-TH/2004-061 \\ TTP04-07 \\}}
 
\title{Radiative return at $\boldsymbol{\Phi}$- and 
$\boldsymbol{B}$-factories: \\
FSR for muon pair production at next-to-leading order~\thanks{Work 
supported in part by BMBF under grant number 05HT9VKB0,
EC 5th Framework Programme under contract HPRN-CT-2002-00311 
(EURIDICE network), TARI project HPRI-CT-1999-00088,
Polish State Committee for Scientific Research (KBN)
under contract 2 P03B 017 24, BFM2002-00568,
Generalitat Valenciana under grant GRUPOS03/013, 
and MCyT under grant FPA-2001-3031.}}

\author{Henryk Czy\.z\inst{1}
\thanks{\email{czyz@us.edu.pl}} 
\and
Agnieszka Grzeli{\'n}ska\inst{1}
\thanks{\email{agrzel@us.edu.pl}}
\and
Johann H. K\"uhn\inst{2}
\thanks{\email{johann.kuehn@uni-karlsruhe.de}}
\and
Germ\'an Rodrigo\inst{3,4}
\thanks{\email{german.rodrigo@cern.ch}}
}

\institute{
Institute of Physics, University of Silesia, PL-40007 Katowice, Poland. 
\and
Institut f\"ur Theoretische Teilchenphysik,
Universit\"at Karlsruhe, D-76128 Karlsruhe, Germany.
\and
Department of Physics, Theory Division, CERN, CH-1211 Geneva 23, Switzerland.
\and
Instituto de F\'{\i}sica Corpuscular, E-46071 Valencia, Spain.
}

\date{April 7, 2004}

\abstract{Muon pair production through the radiative return is of importance for a
measurement of the hadronic production cross section in two ways: it provides
an independent calibration and it may give rise to an important background for
a measurement of the pion form factor. With this motivation the Monte~Carlo
event generator PHOKHARA is extended to include next-to-leading order radiative
corrections to the reaction $e^+e^-\to \mu^+\mu^-\gamma$. Furthermore, virtual
ISR corrections to FSR from pions are introduced, which extends the
applicability of the generator into a new kinematical regime. Finally, the
effect of photon vacuum polarization is introduced into this new version of
the generator.}

\begin{document}
\authorrunning{H. Czy\.z et al. }
\titlerunning{Radiative return at $\boldsymbol{\Phi}$- and 
$\boldsymbol{B}$-factories: FSR for muon pair production at next-to-leading order}

\maketitle

\section{Introduction}
The high luminosity of currently operating $\Phi$- and $B$-meson factories
offers the unique opportunity to measure the famous ratio $R\equiv
\sigma(e^+e^-\to \mathrm{hadrons})/\sigma_{\rm point}$ over a wide range of 
energies in one single experiment, through the radiative return. 
This possibility was suggested a long time ago for pion-pair production~\cite{Zerwas}. 
In view of the capabilities of the currently operating machines, the idea 
was revived in~\cite{Binner:1999bt}, where a Monte~Carlo event 
generator EVA was constructed, which simulates the production of a pion pair
plus one hard photon from initial- and final-state radiation (ISR and FSR). 
The program includes additional collinear radiation from the incoming electrons through
structure function techniques and was used to demonstrate the feasibility of
such a measurement in ongoing experiments.

The generator was subsequently extended to simulate the production of other
hadronic final states: four pions in~\cite{Czyz:2000wh,Czyz:2002np}
and nucleon pairs in~\cite{Nowak},
as well as muons~\cite{Rodrigo:2001kf}. 
The influence of collinear lepton pair radiation on these measurements was
investigated in \cite{Ustron_03_N}.
 
An important step forward was the evaluation of the complete one-loop corrections to
ISR~\cite{Rodrigo:2001jr,Kuhn:2002xg} and the construction of the 
event generator PHOKHARA~\cite{Rodrigo:2001kf}, which included these virtual 
corrections and the corresponding radiation of two photons. 
PHOKHARA, like EVA, can be easily used to simulate reactions
with arbitrary hadronic final states, once a specific model has been chosen
for the relevant matrix element of the hadronic current; it is currently
available for $2\pi$, $4\pi$, $p\bar p$ and $n \bar n$ production 
\cite{Rodrigo:2001kf,Czyz:2002np,Nowak} and of course for muon pairs. 
A number of cross section measurements based on the radiative return, 
which have made use of simulations based on EVA and PHOKHARA,
have been presented recently: a precise measurement of the pion
form factor made by the KLOE Collaboration 
\cite{KLOE:2003,DiFalco:2003xq,Achim_Hadron_03,KLOE:2003_Alghero};
preliminary results on four-prong final states have been obtained 
by the BABAR Collaboration~\cite{babar}.
Extensive discussions of theoretical and experimental aspects of the 
radiative return, including various topics not
covered by the present paper, can be found in the 
literature~\cite{Rodrigo:2001cc,Kuhn:2001,Rodrigo:2002hk,Czyz:PH03,Ustron_03,Alghero_03,CDKMV2000,Adinolfi:2000fv,Aloisio:2001xq,Denig:2001ra,Barbara:Morion,Berger:2002mg,Venanzoni:2002,Blinov,Juliet,Denig:2004ga};
considerations related to the radiative return can 
be found in~\cite{Gluza:eta,Gluza:rad};
perspectives of its use outside hadronic cross section measurements
and the long term potential of the method are outlined in~\cite{Pisa_03}.

Considering measurements with a precision close to one percent, FSR starts 
to become an issue, in particular the contribution from the two-step process 
$e^+e^-\to \gamma \gamma^* (\to \pi^+\pi^-\gamma)$~\cite{Czyz:PH03}.
FSR has to be modeled separately for every different mode.
In leading order (LO), it is now included in PHOKHARA for the $\pi^+\pi^-$ 
and the $\mu^+\mu^-$ modes, in next-to-leading order (NLO), i.e. for the two-step
process with one hard photon from ISR and real plus virtual radiation from the
final state for $\pi^+\pi^-$ only.

It is the purpose of the present paper to extend the same approach to final
states with muon pairs. This is motivated, on the one hand, by the fact that this 
reaction is an important background for the pion form factor measurement, on the
other hand, that it may be used for a measurement of the ratio between hadronic
final states and muon pairs of the same invariant mass. This might
lead to a direct determination of the $R$-ratio, with the cancellation of many
uncertainties that arise in the absolute cross section determination.

Although, technically speaking, the treatment of muons is quite similar to that
of pions, a number of physics aspects are quite different. First of all,
in contrast to radiation from muons, the amplitude for photon radiation from
pions can only be parametrized by a model form factor, and the model
dependence becomes particularly relevant for energetic photons. Then, 
at larger cms energies, relevant e.g. to the $B$-factories, the
leading order FSR with only  one photon in the final state is strongly
suppressed in the pion case, as a consequence of the pion form factor, whereas
the point-like behaviour of the muon leads to important contributions from ISR
as well as FSR in leading order.

In order to arrive at a full result for muon pair production at NLO, 
corrections with combined virtual emission from the initial state and
real hard emission from the final state (Fig.~1c) must be included, and
combined with those from ``simultaneous'' soft real ISR and hard FSR
(Fig.~1a). To offer a uniform program, the same contributions will also be
included for pion pair production. However, we will demonstrate that their 
effect is small already for KLOE energies, in particular for the standard 
KLOE cuts, and completely irrelevant for higher energies.

Finally we study the influence of the leptonic and ha\-dro\-nic vacuum polarization 
on the radiative return, where we adopt the parametrization suggested 
in~\cite{Jeg_web}. For the measurement of the integrated cross section 
with cuts suppressing FSR, vacuum polarization just leads to a multiplicative 
factor that can be taken into account by correcting the result at the end of
the experimental analysis. The forward--backward or the charge asymmetry,
however, is affected by vacuum polarization effects in a non-trivial manner.

Let us briefly outline the content of this paper. The amplitudes for real and
virtual radiation in NLO for muon pair production, which are
the main additional ingredients in the new  program, 
will be introduced in Section 2. Tests of the technical precision of the 
program and some of its physics results will be
presented in Section 3. The ISR correction to real hard FSR for the
$\pi^+\pi^-$ mode will be introduced in Section 4, which also contains a
discussion of the influence of this additional term on present 
experimental studies. The implementation of vacuum polarization and its 
effect on the cross section and asymmetries will be the subject of Section 5. 
Section 6 contains a brief summary and our conclusions.

\section{FSR for muons at next-to-leading order and its implementation in PHOKHARA}

\begin{figure}
\begin{center}
\epsfig{file=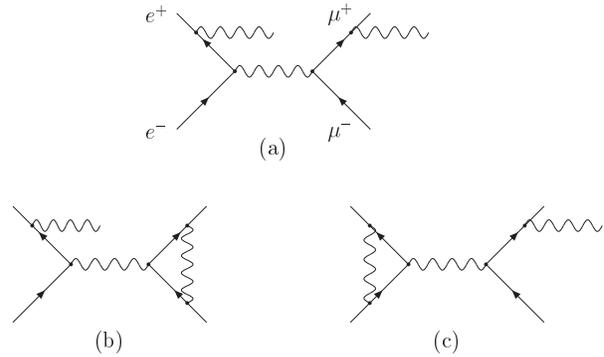,width=8.5cm} 
\caption{NLO corrections to the reaction
$e^+e^-\to\mu^+\mu^-\gamma$: (a) double real emission,
(b) virtual corrections to final-state vertex, and 
(c) virtual corrections to initial-state vertex. 
Only representative Feynman diagrams are depicted.}
\label{mu_diagram}
\end{center}
\end{figure}

The most relevant contributions to FSR at NLO, 
which are depicted schematically in Fig.~\ref{mu_diagram}, 
consist of double real photon emission diagrams, where one photon is 
emitted off the initial-state leptons and the other is emitted from 
the final state (Fig.~\ref{mu_diagram}a), final-state vertex corrections 
to single initial-state photon emission (Fig.~\ref{mu_diagram}b),
and initial-state vertex corrections to single final-state photon 
emission (Fig.~\ref{mu_diagram}c). 

The amplitude for the process depicted in Fig.~\ref{mu_diagram}a
\bea
&&\kern-5pt 
 e^+(p_1,\lambda_{e^+}) \ + \ e^-(p_2,\lambda_{e^-}) \to \nonumber \\
&& \kern-5pt
 \mu^+(q_1,\lambda_{\mu^+}) \ + \ \mu^-(q_2,\lambda_{\mu^-}) \ + \ \gamma(k_1,\lambda_1)
  \ + \ \gamma(k_2,\lambda_2)~,\nonumber \\
\label{eeppgg}
\eea
is given by 
\begin{align}
&M^\mathrm{H}_\mathrm{IFSNLO}(\lambda_{e^+},\lambda_{e^-},
\lambda_{\mu^+},\lambda_{\mu^-},\lambda_1,\lambda_2) =
 -\frac{(4\pi\alpha)^{3/2}}{\hat Q^{2}} \biggl\{ \non \\ &
 v_I^{\dagger}(p_1,\lambda_{e^+}) \
 A(\lambda_{\mu^+},\lambda_{\mu^-},\lambda_1,\lambda_2)
 \ u_I(p_2,\lambda_{e^-}) \non \\&
+v_{II}^{\dagger}(p_1,\lambda_{e^+}) \
 B(\lambda_{\mu^+},\lambda_{\mu^-},\lambda_1,\lambda_2)
 \ u_{II}(p_2,\lambda_{e^-}) 
\biggl\} \non \\ &
 + \ \ \  \bigl( k_1 \leftrightarrow k_2, \lambda_1
 \leftrightarrow \lambda_2 \bigr)~,
\label{hard}
\end{align}
where the index H indicates that both photons are hard,  
\begin{align}
&A(\lambda_{\mu^+},\lambda_{\mu^-},\lambda_1,\lambda_2)  = \non \\ &  
\frac{\left(\varepsilon^*(k_1,\lambda_1)^- k_1^+ 
 - 2\varepsilon^*(k_1,\lambda_1)\cdot p_1\right)
D^{-}(\lambda_2,\lambda_{\mu^+},\lambda_{\mu^-}) } {2 k_1 \cdot p_1}
\non \\ &
 +  \frac{D^{-}(\lambda_2,\lambda_{\mu^+},\lambda_{\mu^-}) 
\left(2\varepsilon^*(k_1,\lambda_1)\cdot p_2 
 -  k_1^+ \varepsilon^*(k_1,\lambda_1)^-\right)}{2 k_1 \cdot p_2}~,
\end{align}
and
\begin{align}
&B(\lambda_{\mu^+},\lambda_{\mu^-},\lambda_1,\lambda_2)  = \non \\ &  
\frac{\left(\varepsilon^*(k_1,\lambda_1)^+ k_1^- 
 - 2\varepsilon^*(k_1,\lambda_1)\cdot p_1\right)
D^{+}(\lambda_2,\lambda_{\mu^+},\lambda_{\mu^-}) } {2 k_1 \cdot p_1}
\non \\ &
 +  \frac{D^{+}(\lambda_2,\lambda_{\mu^+},\lambda_{\mu^-}) 
\left(2\varepsilon^*(k_1,\lambda_1)\cdot p_2 
 -  k_1^- \varepsilon^*(k_1,\lambda_1)^+\right)}{2 k_1 \cdot p_2}~ ,
\end{align}
with 
\begin{align}
\hat Q = p_1 + p_2 - k_1 = q_1 + q_2 + k_2~, \qquad s' = \hat Q^2~.
 \end{align}
\(D^\mu\) is the current describing the \(\mu^+ \mu^-\gamma\) 
final state \cite{Czyz:2002np}, which we report below for completeness
\begin{align}
D^\mu(\lambda_2,\lambda_{\mu^+},\lambda_{\mu^-})& =
 i e\biggl\{ 
 u_I^{\dagger}(q_2,\lambda_{\mu^-}) 
 \tilde{A}^{\mu}(\lambda_2)  v_I(q_1,\lambda_{\mu^+}) \non \\ &
+u_{II}^{\dagger}(q_2,\lambda_{\mu^-}) 
 \tilde{B}^{\mu}(\lambda_2)  v_{II}(q_1,\lambda_{\mu^+})\biggr\} ,
\end{align}
with
\begin{align}
\tilde{A}^{\mu}(\lambda_2)&  = 
\frac{\left( 2 q_2 \cdot \varepsilon^*(k_2,\lambda_2) 
+ \varepsilon^*(k_2,\lambda_2)^-k_2^+\right)\sigma^{\mu-}}
{2 k_2 \cdot q_2} \non \\ &
 -  \frac{\sigma^{\mu-} 
\left(2 q_1 \cdot \varepsilon^*(k_2,\lambda_2) 
 +  k_2^+ \varepsilon^*(k_2,\lambda_2)^-\right)}{2 k_2 \cdot q_1}~ ,
\end{align}
\begin{align}
\tilde{B}^{\mu}(\lambda_2)& 
 = \frac{\left( 2 q_2 \cdot \varepsilon^*(k_2,\lambda_2) 
+ \varepsilon^*(k_2,\lambda_2)^+k_2^-\right)\sigma^{\mu+}}
{2 k_2 \cdot q_2} \non \\ &
 -  \frac{\sigma^{\mu+} 
\left(2 q_1 \cdot \varepsilon^*(k_2,\lambda_2) 
 +  k_2^- \varepsilon^*(k_2,\lambda_2)^+\right)}{2 k_2 \cdot q_1}~ ,
\end{align}
and $\sigma^{\mu\pm}=(I,\pm\sigma_i)$, where $\sigma_i$, $i=1,2,3$, are the 
Pauli matrices and $a^\pm = a^\mu\sigma^\pm_\mu $ for any four-vector $a^\mu$.

The virtual (Fig.~\ref{mu_diagram}b) plus soft photon corrections to the 
final-state vertex can be written as~\cite{Berends:1973np}
\bea
d\sigma^\mathrm{V+S}_\mathrm{IFSNLO,f} = \frac{\alpha}{\pi} \ 
   \eta^\mathrm{V+S}(s', E_2^\mathrm{cut}) \ 
   d\sigma^{(0)}_{\mathrm{ISR}}(s')~,
\label{sig_VS}
\eea
where $d\sigma^{(0)}_{\mathrm{ISR}}$ is the leading order 
$e^+e^- \to \mu^+\mu^-\gamma$ cross section, with the photon 
emitted off the initial leptons only, and
\begin{align}
&\eta^{\mathrm{V+S}}(s', E^{\mathrm{cut}}_{2}) = -2\biggl[\frac{1+\beta_{\mu}^2}
{2\beta_{\mu}} \log(t) + 1 \biggr] \log(2w')  \nonumber \\ &
-\frac{\log(t)}{s'\beta_{\mu}}\biggl[\frac{5}{2}s'-7m_{\mu}^2 
+3m_{\mu}^2x_{\mu}\biggr] - 2 + 
\log\biggl(\frac{1-\beta_{\mu}^2}{4}\biggr) \nonumber \\ &
-\frac{1+\beta_{\mu}^2}{\beta_{\mu}}\biggl[
2\Li_2 \biggl( \frac{2\beta_{\mu}}{1+\beta_{\mu}} \biggr) - \frac{\pi^2}{2} 
-\log(t)\log\biggl(\frac{1+\beta_{\mu}}{2}\biggr)\biggl]~. \nonumber  \\
\label{etavs}
\end{align}
$ E^{\mathrm{cut}}_{2}$ is the maximal energy of the soft 
photon in the $s'$ rest frame, and
\bea
\beta_{\mu} = \sqrt{1 - 4m_{\mu}^{2}/s'}~, 
\qquad t = \frac{1-\beta_{\mu}}{1+\beta_{\mu}}~, \nonumber \\
w' =  E_{2}^{\mathrm{cut}}/\sqrt{s'}~,\qquad
\frac{1}{x_{\mu}} = \frac{s'}{2m_{\mu}^2}+1~.
\label{beta}
\eea
The virtual  (Fig.~\ref{mu_diagram}c) plus soft photon corrections 
to the initial-state vertex can be written as~\cite{Berends:1988np}
\bea
d\sigma^\mathrm{V+S}_\mathrm{IFSNLO,i} = \frac{\alpha}{\pi} \ 
   \delta^{\mathrm{V+S}}(s,E_\gamma^\mathrm{min}) \ 
   d\sigma^{(0)}_{\mathrm{FSR}}(s)~,
\label{sigm_i}
\eea
where $d\sigma^{(0)}_{\mathrm{FSR}}$ is the leading order 
$e^+e^- \to \mu^+\mu^-\gamma$ cross section, with the photon 
emitted off the final leptons only, and, for $m_e^2\ll s$:
\begin{align}
\delta^{\mathrm{V+S}} = 2 \biggl\{ 
(L-1)\log{(2w)}+\frac{3}{4}L-1+\zeta(2)\biggr\}~.
\label{del_i}
\end{align}
$E_\gamma^\mathrm{min} $ is the maximal energy of the soft 
photon (or the minimal energy of the hard photon) in the 
$s$ rest frame, while 
\begin{align}
L = \log \biggl( \frac{s}{m_{e}^2} \biggr) \ {\rm and} \ \
 w = E_\gamma^\mathrm{min}/\sqrt{s}~.
\end{align}
To match hard, soft and virtual radiation smoothly, the energy cutoff 
(\( E_2^\mathrm{cut}\)) has to be transformed from the rest frame of the 
$\mu^+ \mu^- \gamma$  system (with the photon emitted from the final state)  
to the laboratory frame (\(e^+e^-\) cms frame) (\(E_\gamma^\mathrm{min} \)).
In fact it is necessary to recalculate the soft photon contribution,
as the cut on \(Q^2=(q_1+q_2)^2\) depends in the latter case on the angle 
between the two emitted photons. Now
\begin{align}
&\eta^{\mathrm{V+S}}(s',E^{\mathrm{cut}}_{2}) = -2\biggl[\frac{1+\beta_{\mu}^2}
{2\beta_{\mu}} \log(t) + 1 \biggr] \nonumber \\ & 
\times  \biggl[ \log(2w) + 1 + \frac{s'}{s'-s} 
\log\biggl(\frac{s}{s'} \biggr) \biggr]- 2 + 
\log\biggl(\frac{1-\beta_{\mu}^2}{4}\biggr) \nonumber \\&
-\frac{\log(t)}{s'\beta_{\mu}}\biggl[\frac{5}{2}s'-7m_{\mu}^2 
+3m_{\mu}^2x_{\mu}\biggr] -\frac{1+\beta_{\mu}^2}{\beta_{\mu}} \biggl[
2\Li_2 \biggl( \frac{2\beta_{\mu}}{1+\beta_{\mu}} \biggr)\nonumber \\ &
 - \frac{\pi^2}{2} 
-\log(t)\log\biggl(\frac{1+\beta_{\mu}}{2}\biggr)\biggl] \ .
\label{etavsp}
\end{align}
Using Eqs.~(\ref{hard}) to (\ref{etavsp}) the implementation of FSR in 
combination with ISR is straightforward.

We have not included diagrams where two photons are emitted from the final
state, neither final-state vertex corrections with associated real radiation 
from the final state \cite{KKMC}.
This constitutes a radiative correction to FSR and will give 
non negligible contributions only for those cases, where at least one 
photon is collinear with one of the muons.
Box diagrams with associated real radiation from the initial- or the 
final-state leptons, as well as pentagon diagrams, are also neglected.
As long as one considers charge symmetric
observables only, their contribution is neither divergent in the soft nor
the collinear limit and thus of order $\alpha/\pi$ without any enhancement
factor. We want to stress that we have included only C-even gauge 
invariant sets of diagrams (see Fig.~9 of Ref.~\cite{Czyz:PH03} for a 
graphical representation of the equivalent set of diagrams in pion pair 
production). The box and pentagon diagrams that we have neglected 
are related to ISR-FSR interferences, and therefore are suppressed 
after suitable kinematical cuts used within the radiative return method. 
One could also neglect some small contributions from the diagrams taken
into account, but they are kept for the sake of completeness, as
the full 1-loop radiative corrections will be consider in a future
publication. At this point these small contributions will become 
relevant allowing accurate calculations not restricted to radiative 
return physical configurations.

\section{Tests of the Monte Carlo program and discussion of physical 
results} 

\begin{figure}
\begin{center}
\epsfig{file=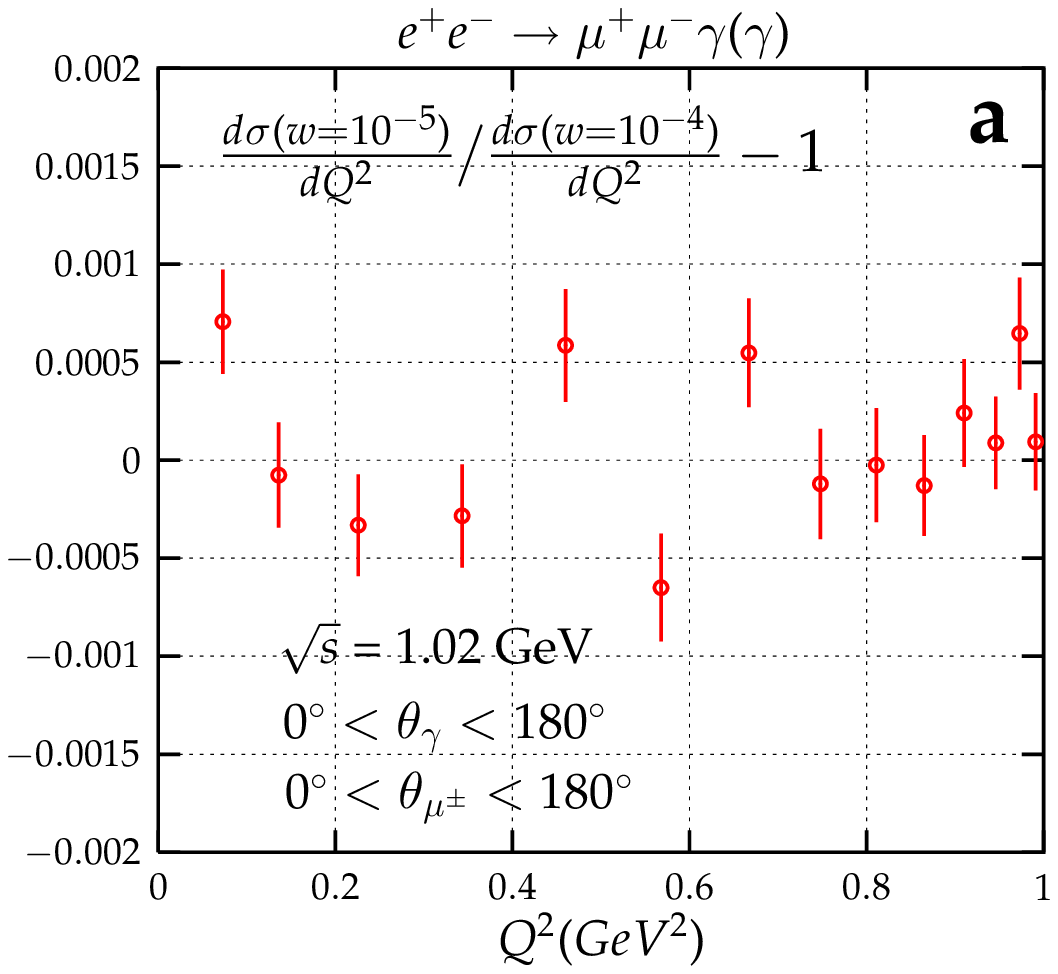,height=7cm,width=8.5cm} 
\epsfig{file=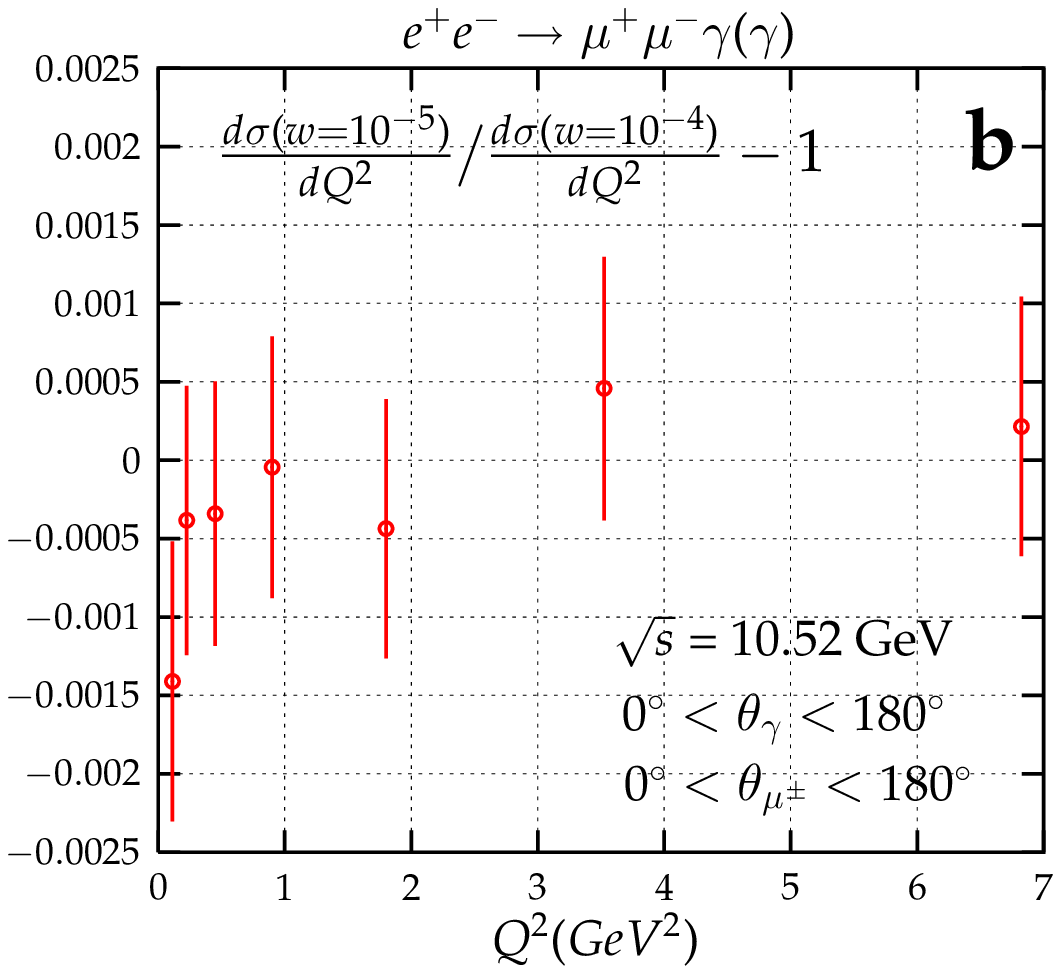,height=7cm,width=8.5cm} 
\caption{Comparison of the $Q^2$ distribution for two
values of the soft photon cutoff ($w = 10^{-4}$ vs. $10^{-5}$)
for $\sqrt{s}=1.02$~GeV (a) and $\sqrt{s}=10.52$~GeV (b). 
One of the photons was required to have energy $>$ 10~MeV 
(for $\sqrt{s}=1.02$~GeV) and $>$ 100~MeV (for $\sqrt{s}=10.52$~GeV).
One sigma statistical errors are shown for the ratio of the 
two distributions.
\label{fig:1.02_mu_-5-4}}
\end{center}
\end{figure}

\begin{table}
\begin{center}
\caption{Total cross section (nb), corresponding to the $Q^2$ 
distributions from Fig.~\ref{fig:1.02_mu_-5-4},
for the process $e^+ e^- \rightarrow \mu^+ \mu^- \gamma$
at NLO, for different values of the soft photon cutoff.}
\label{tab:epstest}
\begin{tabular}{ccc}
$w$ & $\sqrt{s}=$ 1.02~GeV  & 10.52~GeV \\ \hline 
$10^{-4}$ & 28.237(1)  & 0.19854(7) \\
$10^{-5}$ & 28.236(1)  & \hskip 0.16 cm 0.19883(14)\\ \hline
\end{tabular}
\end{center}
\end{table}

A number of tests were performed to ensure the technical 
precision of the new version of PHO\-KHA\-RA. The square of the matrix 
element, summed over polarizations of the final particles
and averaged over polarizations of the initial particles, 
was calculated with FORM~\cite{FORM}, using the standard trace method.
External gauge invariance was checked analytically when using the trace
method and numerically for the amplitude calculated with the helicity
am\-pli\-tu\-de me\-thod. The two results for the square of the matrix element
summed over polarizations were compared numerically. The code based
on the result from the trace method was written in quadruple precision
to reduce cancellations. The code based on the result obtained with the 
helicity amplitude method uses double precision for real and complex 
numbers and is now incorporated in the code of PHO\-KHA\-RA 4.0. 
Agreement of 13 significant digits (or better) was found between the 
two codes. The sensitivity of the integrated cross section
to the choice of the cutoff $w$ can be deduced from  
Fig.~\ref{fig:1.02_mu_-5-4} and Table~\ref{tab:epstest}.
For simplicity the same separation parameter $w$ was chosen for 
ISR and FSR corrections. 
Choosing $w=10^{-4}$ or less, the result becomes independent of $w$, 
up to expected systematic differences of order 10$^{-4}$.
These systematic differences are visible already in Fig.~2a,
where $\chi^2$/d.o.f = 30/14, while in Fig.~2b statistical errors
are about 0.1\%, and  $\chi^2$/d.o.f = 3.5/7. 
The tests prove that the analytical formula describing soft photon 
emission as well as the Monte Carlo integration in the soft photon 
region are well implemented in the program. 

For the purpose of further tests we have also calculated
the integrated contribution of hard photon emission from the final 
state to the ISR spectrum. Integrating over all angles and
energies, from $E_2^\mathrm{cut}$ (defined in the $s'$ rest frame)
to the kinematical limit of the final-state photon, we find
that this contribution can be written as
\bea
d\sigma^\mathrm{H}_\mathrm{IFSNLO,f} = \frac{\alpha}{\pi} \ 
   \eta^\mathrm{H}(s', E_2^\mathrm{cut}) \ 
   d\sigma^{(0)}_{\mathrm{ISR}}(s')~,
\label{sighard}
\eea
where
\begin{align}
& \eta^\mathrm{H}(s',E_2^\mathrm{cut}) =
- \frac{1+\beta_{\mu}^2}{\beta_{\mu}}\biggl[\Li_{2}\biggl( 
1-\frac{t_{m}}{t} \biggr) - \Li_{2}(t_{m}t) \nonumber \\&
+ \zeta(2) + \frac{\log^{2}(t)}{2}\biggr] 
- \log(t_{m}-t)\biggl[\frac{1+\beta_{\mu}^2}{\beta_{\mu}}
\log \biggl(\frac{t_{m}}{t} \biggr)  \nonumber \\&
-2 \biggr]+ \frac{\log(t_m)}{\beta_{\mu}}\biggl\{  
\frac{1-\beta_{\mu}^2}{1-\beta_{m}^2}\biggl[ 
 1+\frac{x_{\mu}(2-\beta_{m}^2)}{(1-\beta_{m}^2)}\biggr]
\nonumber \\&
+ (1+\beta_{m}^2)\biggl[ \log\biggl( \frac{4t(\beta_{\mu}^2 - \beta_{m}^2)}
{(1-\beta_{\mu}^2)(1-\beta_{m}^2)} \biggr) + \frac{\log(t_m)}{2} \biggr]
\nonumber \\&
-\frac{1}{8}\biggl[ 17-\beta_{\mu}^2-2x_{\mu}\biggr]
  \biggr\} - \frac{x_{\mu}}{4\beta_{m}\beta_{\mu}} 
\biggl\{33 -17\beta_{\mu}^2  \nonumber \\& 
+\frac{1}{1-\beta_{m}^2}\biggl[ -27+11\beta_{\mu}^2
-\frac{6(1-\beta_{\mu}^2)}{1-\beta_{m}^2}\biggr] \biggr\}  \nonumber \\&
- 2 \log(1-t_mt)~,
\label{etah}
\end{align}
%
$t$, $x_\mu$ and $\beta_\mu$ are defined in \Eq{beta}, and
\bea
\beta_m \equiv \sqrt{1-\frac{4m^2_\mu}{Q_{m}^{2}}}~, \qquad 
t_{m} = \frac{1-\beta_m}{1+\beta_m}~. \nonumber  
\eea
Here \(Q_{m}^{2}\) is the maximum value of \(Q^2\) 
\begin{equation}
Q_{m}^{2} = s' - 2 E_2^{\mathrm{cut}} \sqrt{s'}~.  \nonumber  
\end{equation}

For small $E_2^\mathrm{cut}$, $w = E_2^\mathrm{cut}/\sqrt{s'} \ll 1$,
the function $\eta^\mathrm{H}$ reduces to
\begin{align}
& \eta^\mathrm{H}(s',E_2^\mathrm{cut})  \simeq 
\log(2w) \biggl[2+\frac{1+\beta_{\mu}^{2}}{\beta_{\mu}}\log(t)\biggr] 
\nonumber \\ &
+ \frac{1+\beta_{\mu}^2}{\beta_{\mu}}\biggl[ \Li_2(t^2)
-\zeta(2) -2\log(t)\log\biggl(\frac{1+\beta_{\mu}}{2}\biggr)\biggr]
\nonumber \\ &
+\frac{\log(t)}{8\beta_{\mu}}\biggl[ -5 + \beta_{\mu}^2
+6x_{\mu}\biggr] 
-2\log\biggl(\frac{4\beta_{\mu}^2}{1-\beta_{\mu}^2}  \biggr)\nonumber \\ &
+ \frac{3}{2}x_{\mu} + \frac{11}{4} \ \ .
\label{etahsoft}
\end{align}

\begin{figure}
\begin{center}
\epsfig{file=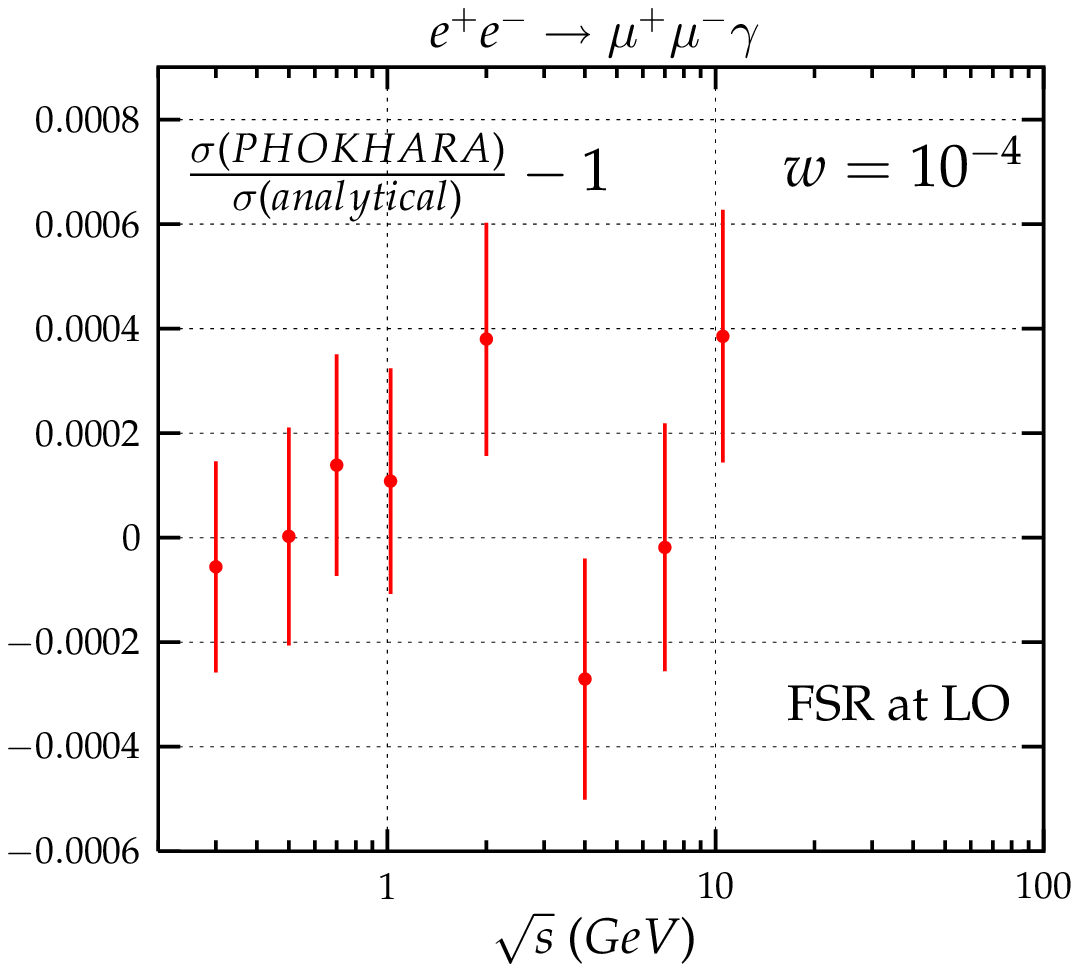,height=7cm,width=8.5cm} 
\caption{Comparison between the LO FSR cross section calculated analytically 
(Eq.~(\ref{sigFSR})) and calculated by PHO\-KHA\-RA for a fixed
value of $w=E_{\gamma}^\mathrm{min}/\sqrt{s}$.  
Errors represent one sigma statistical errors.}
\label{fig:eta_10-4}
\end{center}
\end{figure}

\noindent 
Adding virtual, soft (\Eq{etavs}) and hard (\Eq{etahsoft}) corrections, 
the familiar correction factor~\cite{CKK}
\begin{align}
& \eta(s')  = \frac{1+\beta_{\mu}^2}{\beta_{\mu}} \biggl[ 4\Li_{2}(t) 
+ 2\Li_{2}(-t)
\nonumber \\ &
- \log(t)\log\biggl(\frac{(1+\beta_{\mu})^3}{8\beta_{\mu}^2}
\biggr) \biggr] \nonumber + 3 \log\biggl(\frac{1-\beta_{\mu}^2}{4\beta_{\mu}}
\biggr) - \log(\beta_{\mu}) \nonumber \\ &
+ \frac{2}{\beta_{\mu}(3-\beta_{\mu}^2)} \biggl[
\frac{3\beta_{\mu}}{8}(5-3\beta_{\mu}^2)
 - \frac{3\log(t)}{48}(33+22\beta_{\mu}^2-7\beta_{\mu}^4) \biggr]\nonumber \\ &
\label{etatot}
\end{align}
is recovered.

For single-photon emission from the final state and no further photon 
radiation, a formula similar to Eq.~(\ref{sighard}) holds:
\bea
\sigma^\mathrm{H}_\mathrm{FSR} = \frac{\alpha}{\pi} \ 
   \eta^\mathrm{H}(s, E_\gamma^\mathrm{min}) \ 
   \sigma^{(0)}_{e^+e^-\to\mu^+\mu^-}(s)~,
 \label{sigFSR}
\eea
with $\eta^\mathrm{H}$ defined in Eq.~(\ref{etah}), $E_\gamma^\mathrm{min}$
defined now in the $s$ rest frame and $\sigma^{(0)}_{e^+e^-\to\mu^+\mu^-}$
the lowest order cross section of the process $e^+e^-\to\mu^+\mu^-$.

\begin{figure} 
\begin{center}
\epsfig{file=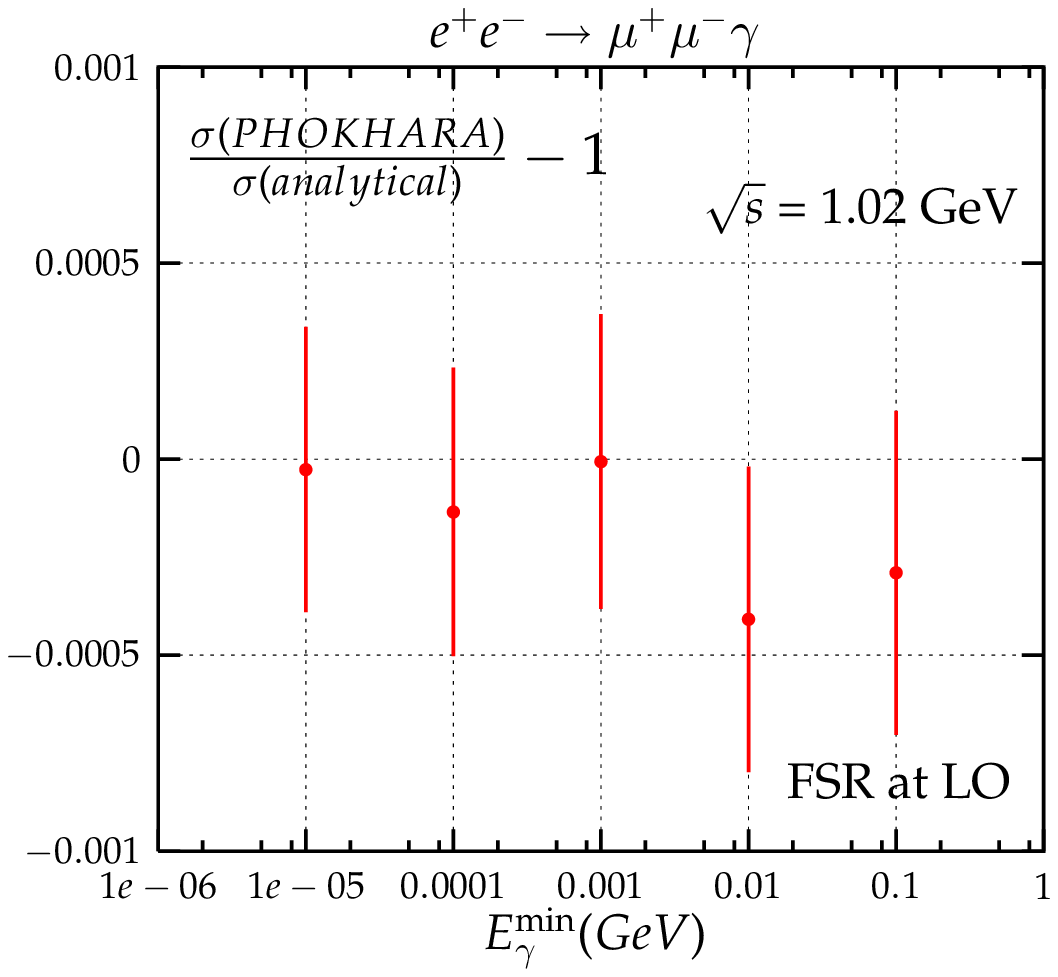,height=7cm,width=8.5cm}\vspace{0.5 cm}
\epsfig{file=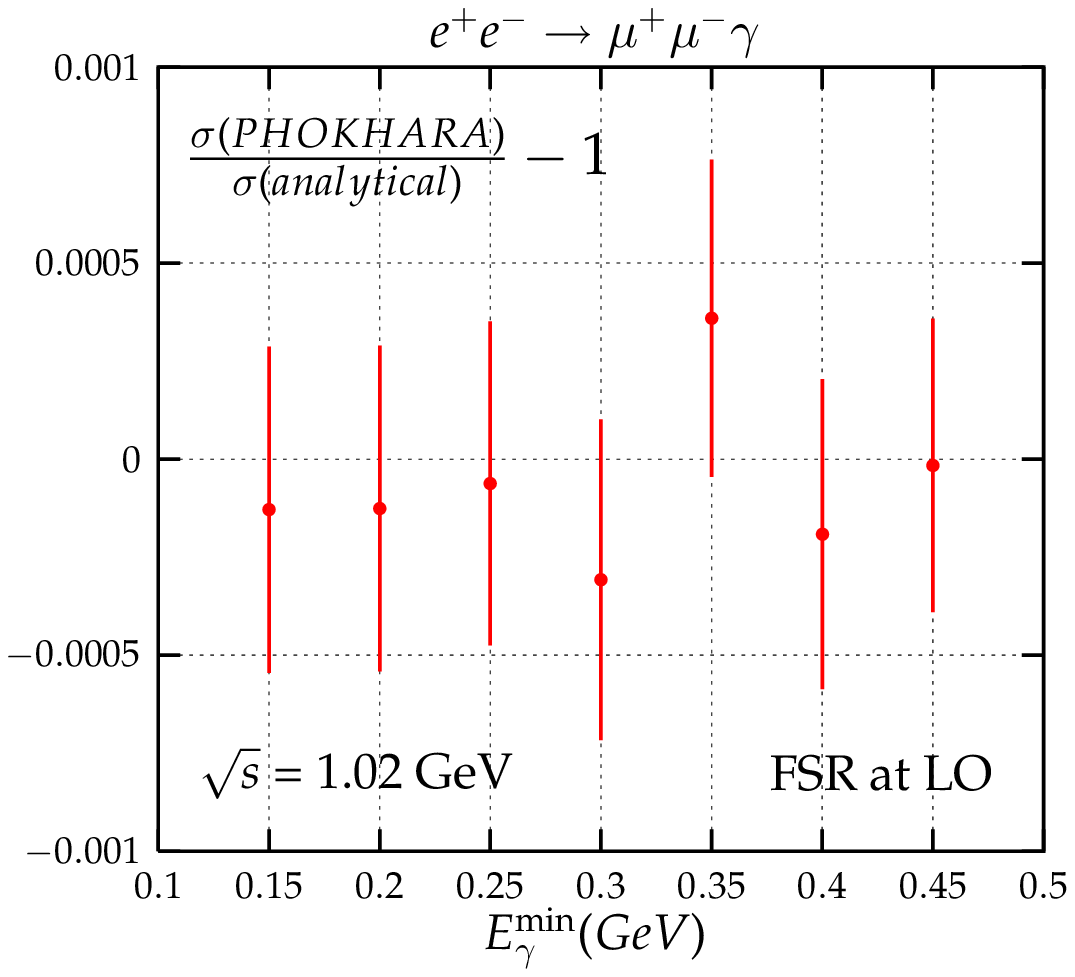,height=7cm,width=8.5cm}
\caption{Comparison between the LO FSR cross section calculated analytically 
(Eq.~(\ref{sigFSR})) and calculated by PHO\-KHA\-RA for a fixed
value of \( \sqrt{s} = 1.02 \)~GeV. Errors represent one sigma statistical 
errors.}
\label{fig:eg_min_1.02}
\end{center}
\end{figure} 

\begin{figure}
\begin{center}
\epsfig{file=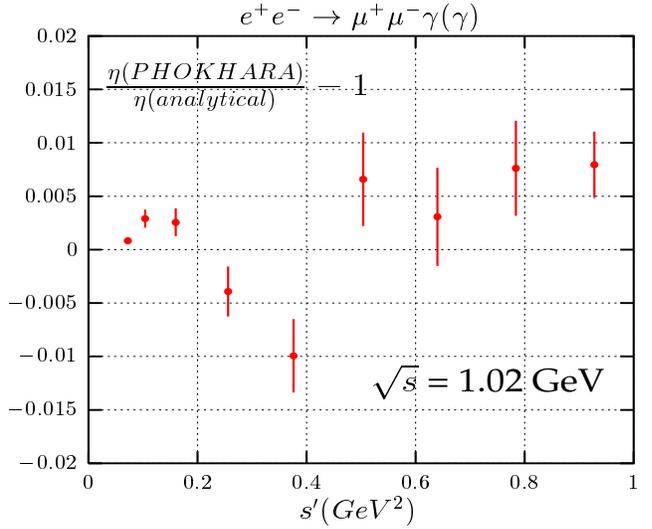,height=7cm,width=8.5cm} 
\caption{Comparison between analytical function \(\eta\) (Eq.~(\ref{etatot}))
and \(\eta\) obtained from PHO\-KHA\-RA for $\sqrt{s} = $ 1.02 GeV.}
\label{fig:eta_2ph}
\end{center}
\end{figure}

The results of the tests are collected in Figs.~\ref{fig:eta_10-4} 
and~\ref{fig:eg_min_1.02} proving the excellent technical precision
of this part of PHO\-KHA\-RA, well below 0.05\%. Results of similar tests 
at NLO are collected in Fig.~\ref{fig:eta_2ph}, where results of PHOKHARA
are compared with the analytical result of Eq.~(\ref{etatot}). 
The latter checks the technical precision of the implementation of 
double-photon emission, where one photon is emitted from the initial state 
and the other from the final state, together with the corresponding virtual 
and soft corrections to the final-state vertex. An agreement of 1\% on 
the function $\eta$ means in fact an agreement better than
$10^{-4}$ on the cross section thanks to the additional factor 
$\alpha/\pi$ (compare Eq.~(\ref{sig_VS}) and Eq.~(\ref{sigFSR})).

\begin{figure}
\hfill\epsfig{file=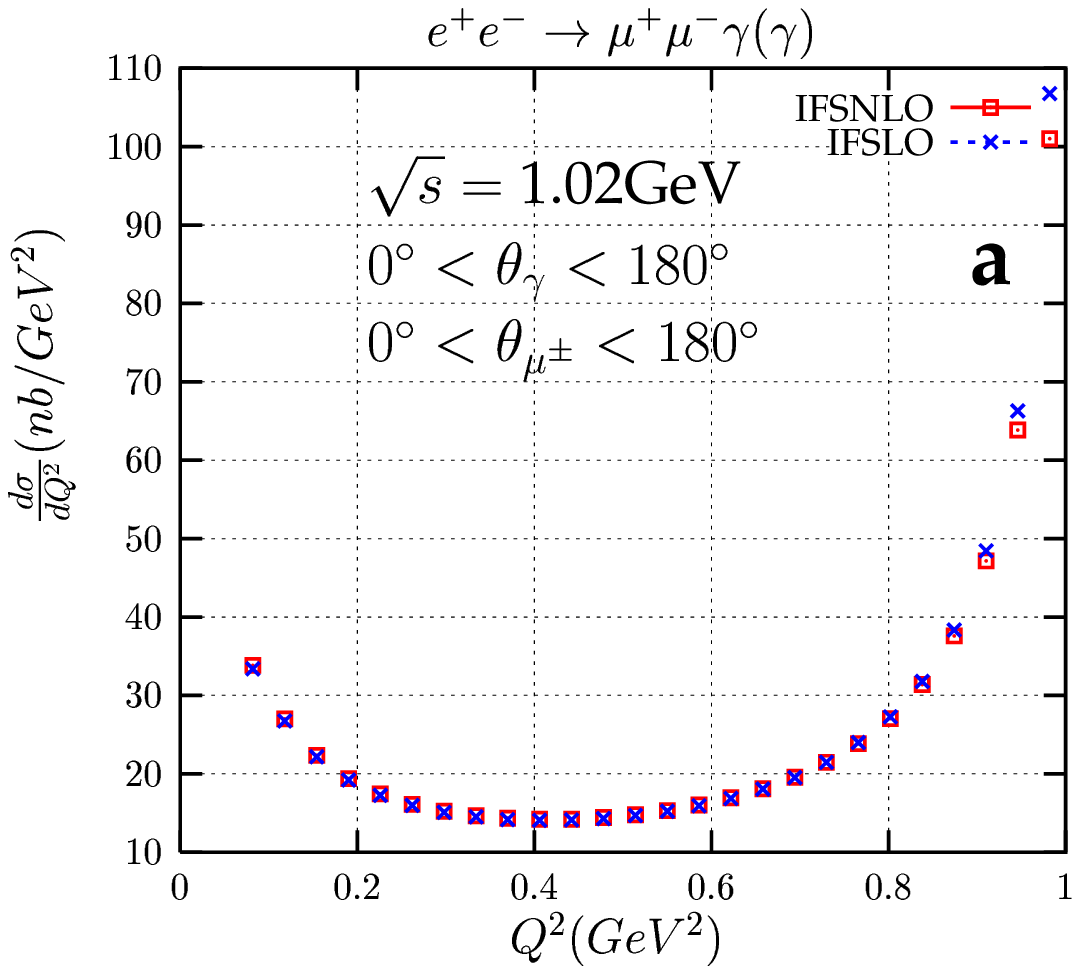,width=7cm} 
\hfill\epsfig{file=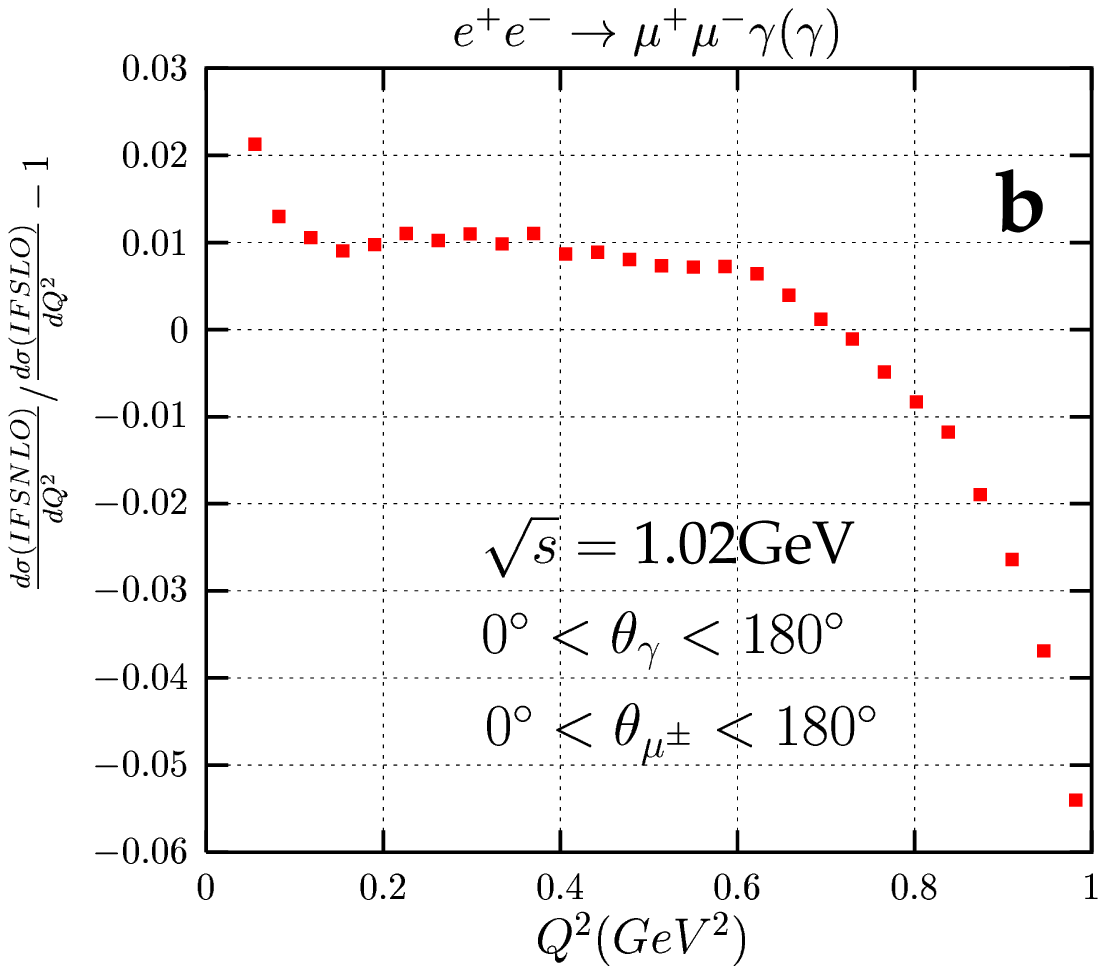,width=7cm} 
\hfill\epsfig{file=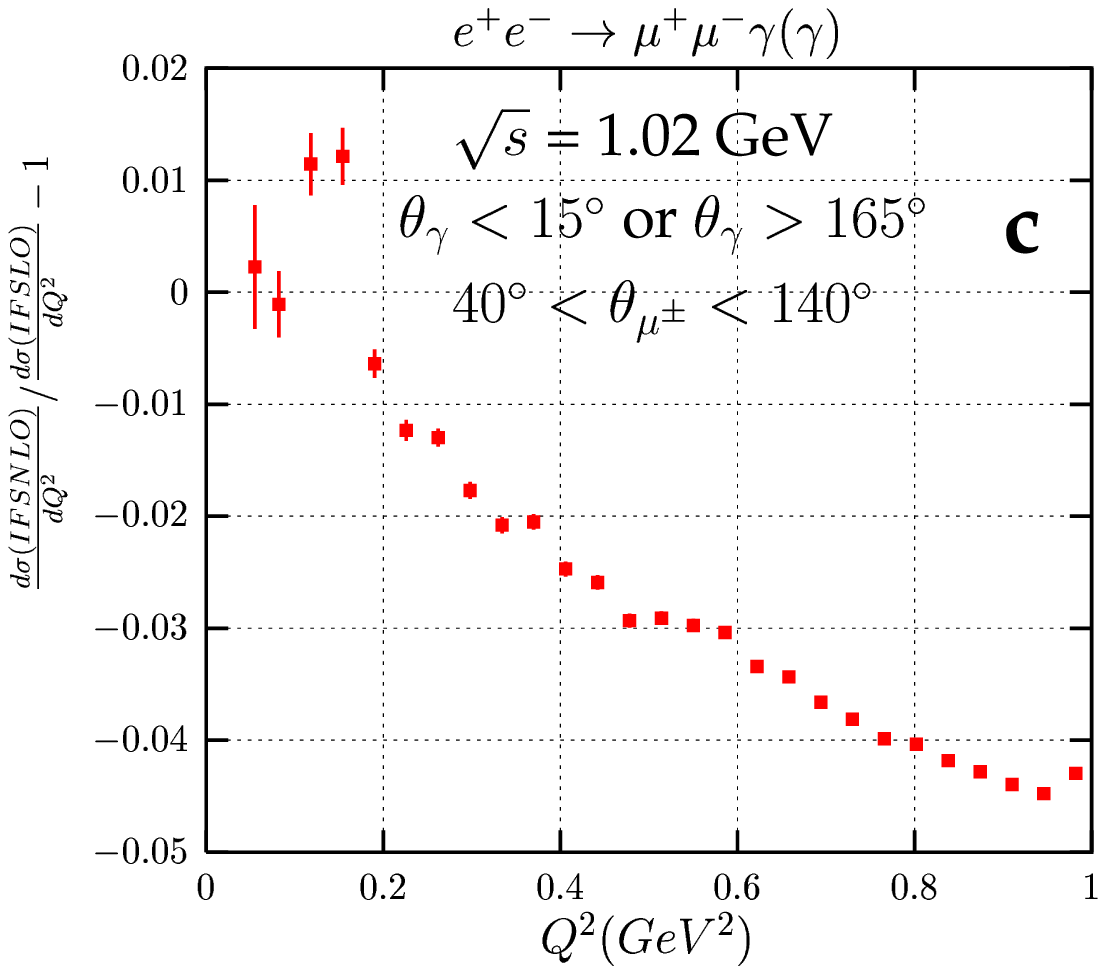,width=7cm} 
\begin{center}
\caption{The size of the NLO FSR corrections (IFSNLO) compared
with the sum of NLO ISR and LO ISR and FSR contributions (IFSLO) to the  
$e^+e^-\to \mu^+\mu^- \gamma(\gamma)$ cross section: 
a) $Q^2$ differential cross sections with no angular cuts;
b) relative difference of the cross sections with no angular cuts;
c) relative difference of the cross sections with angular cuts
chosen to suppress FSR at LO.}
\label{fig:1.02_nlo_lo}
\end{center}
\end{figure}

The relative size the NLO FSR contributions to the 
$e^+e^-\to \mu^+\mu^-\gamma(\gamma) $ cross section introduced in PHO\-KHA\-RA
depends on the event selection used. In the following, we will indicate some
of its characteristic features. In Fig.~\ref{fig:1.02_nlo_lo}a the
$Q^2$ differential cross section is shown with two peaks at low 
(`soft' muon pair production) and large $Q^2$ (soft photon emission) values.
In Fig.~\ref{fig:1.02_nlo_lo}b the corresponding relative NLO FSR contributions
are shown. They are $Q^2$-dependent and might be as big as a few per cent. 
As seen by comparison of Figs.~\ref{fig:1.02_nlo_lo}b and~\ref{fig:1.02_nlo_lo}c 
the relative size of this contribution does depend on the event selection used.

\begin{figure}
\begin{center}
\epsfig{file=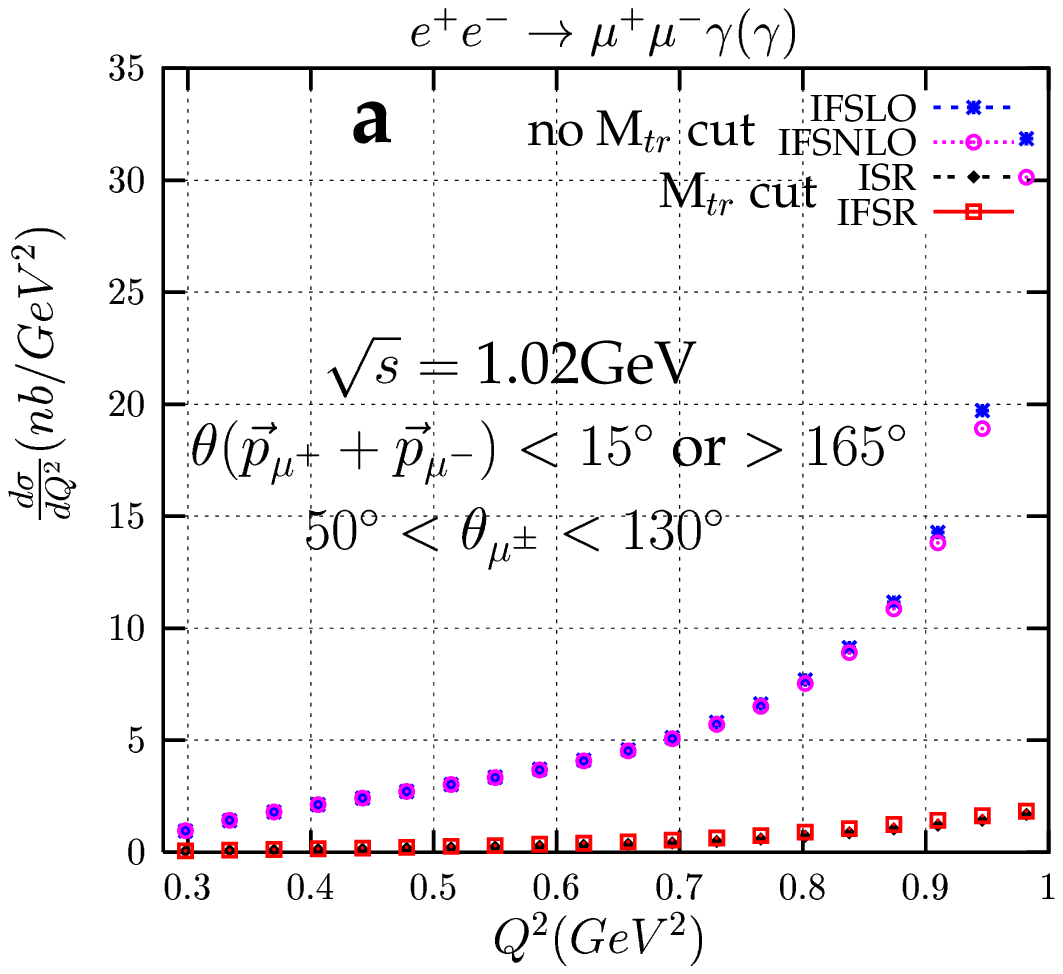,width=8.cm} 
\epsfig{file=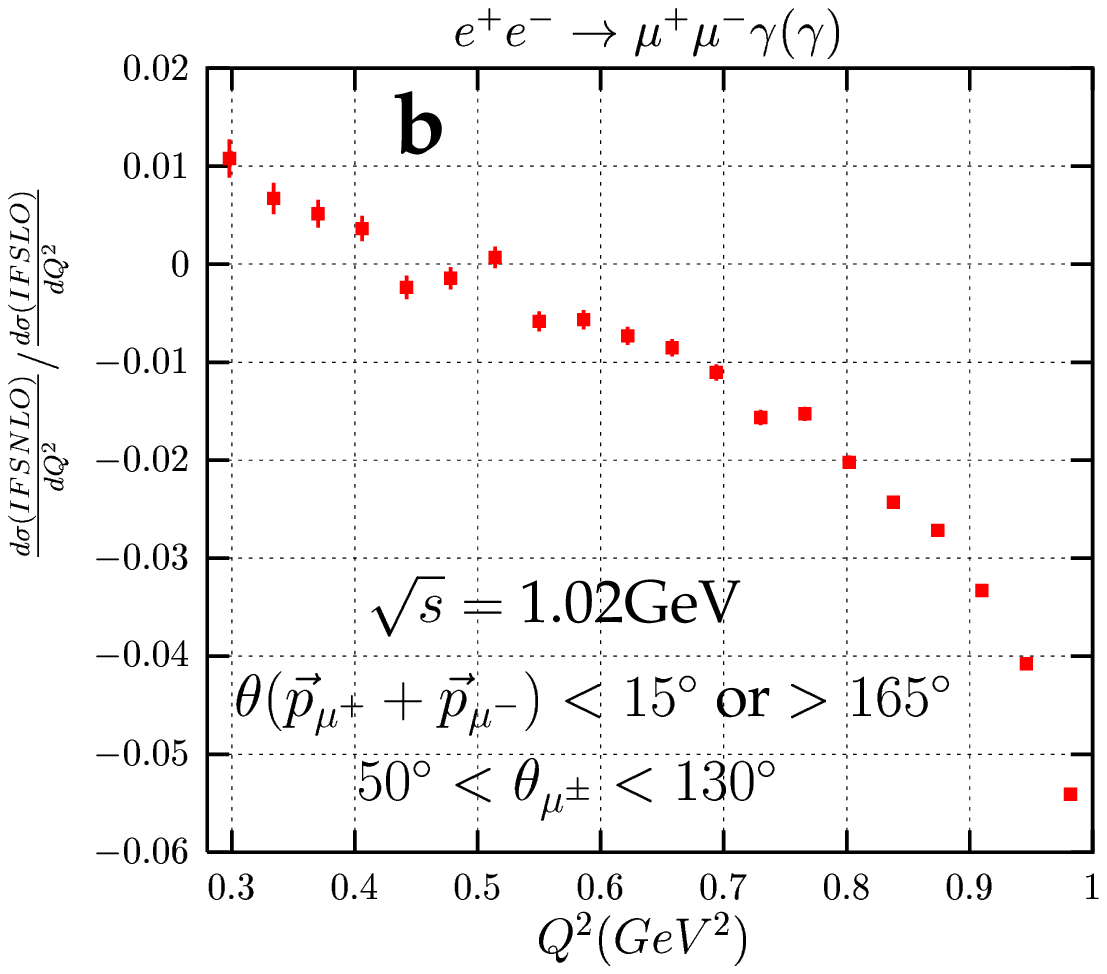,width=8.cm} 
\epsfig{file=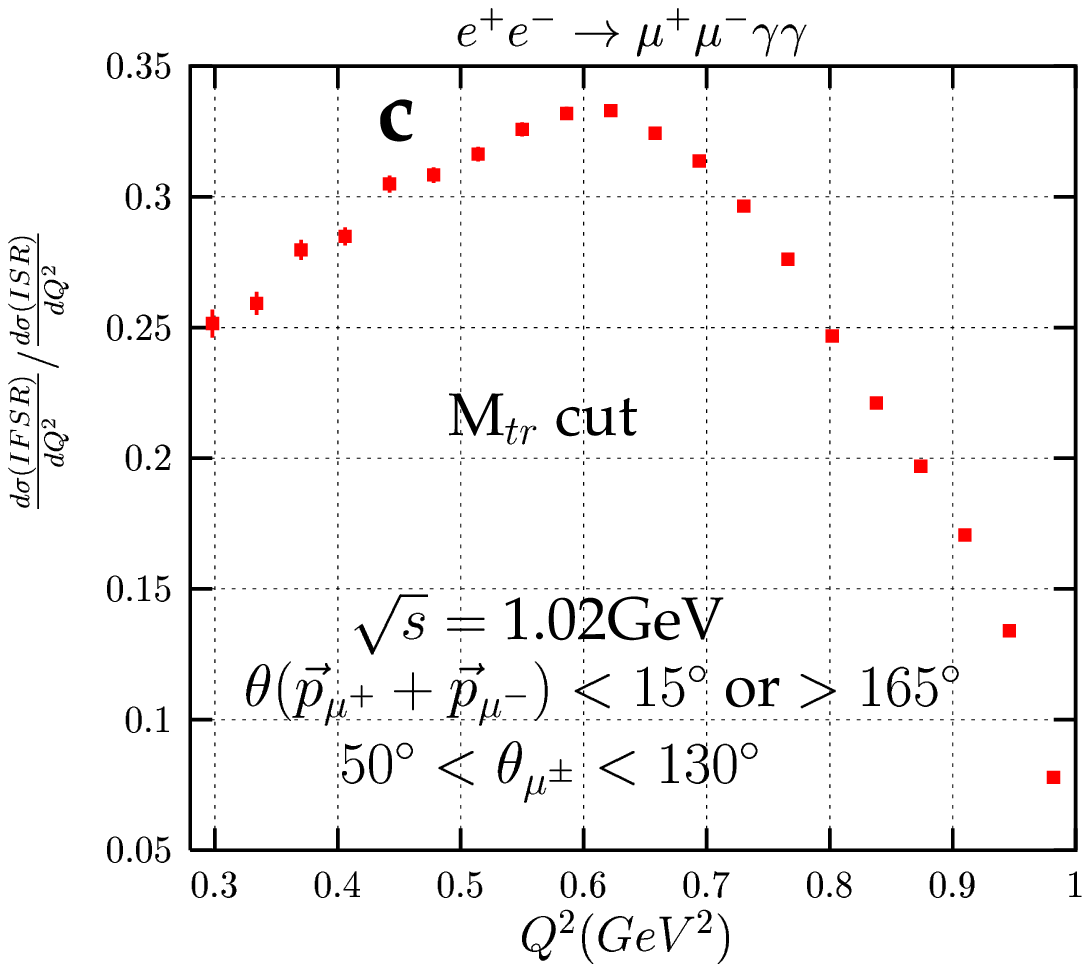,width=8.cm} 
\caption{The size of the NLO FSR corrections (IFSNLO) compared
with the sum of NLO ISR and LO ISR and FSR contributions (IFSLO) to the  
$e^+e^-\to \mu^+\mu^-\gamma(\gamma)$ cross section: 
a) $Q^2$ differential cross sections with angular cuts used by KLOE;
b) relative difference of the cross sections with angular cuts 
used by KLOE (no track mass cut);
c) relative difference of the cross sections with angular cuts
used by KLOE (with track mass cut). The track mass cut allows only
for two-photon events and here IFSR stands for two- and one-photon ISR +
one-photon FSR, while ISR stands for two-photon ISR only.}
\label{fig:ang_trk}
\end{center}
\end{figure}

\begin{figure}
\begin{center}
\epsfig{file=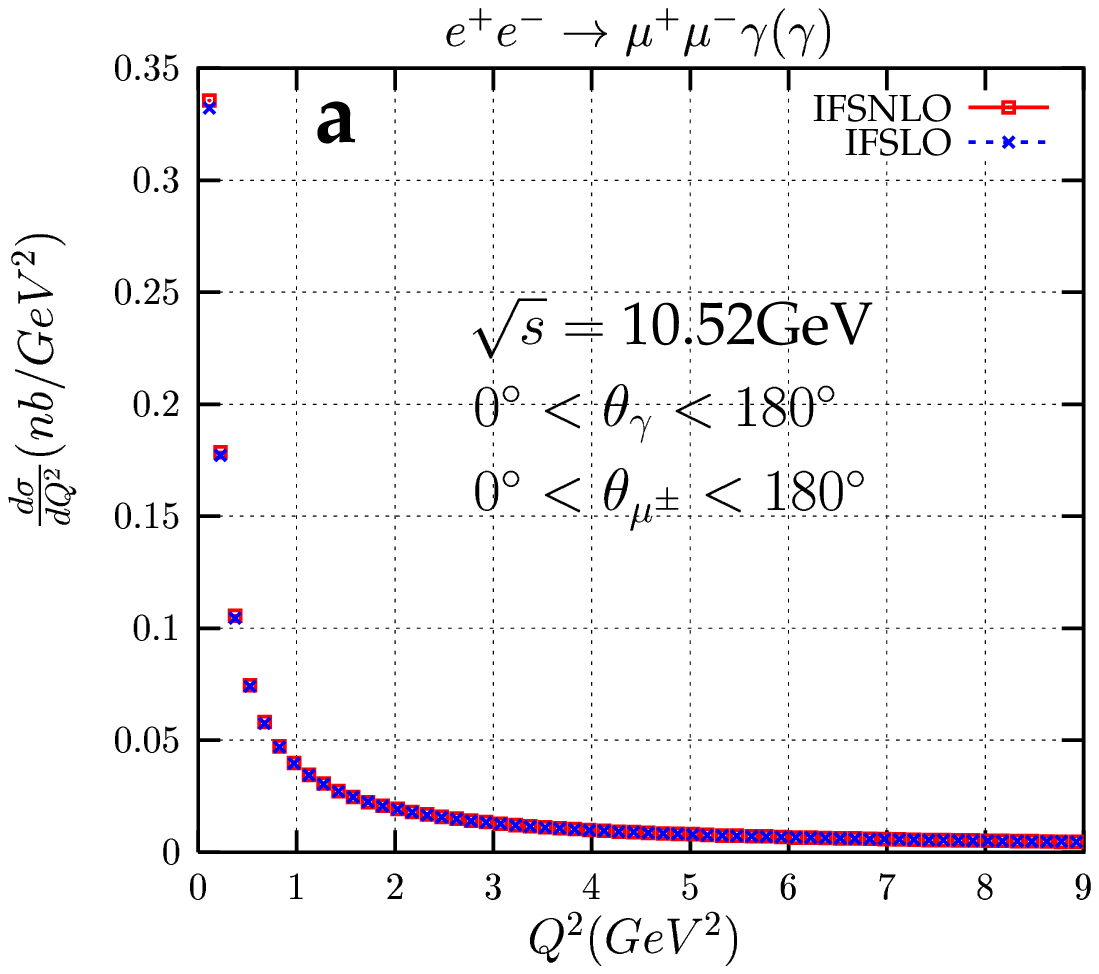,width=8.5cm} 
\epsfig{file=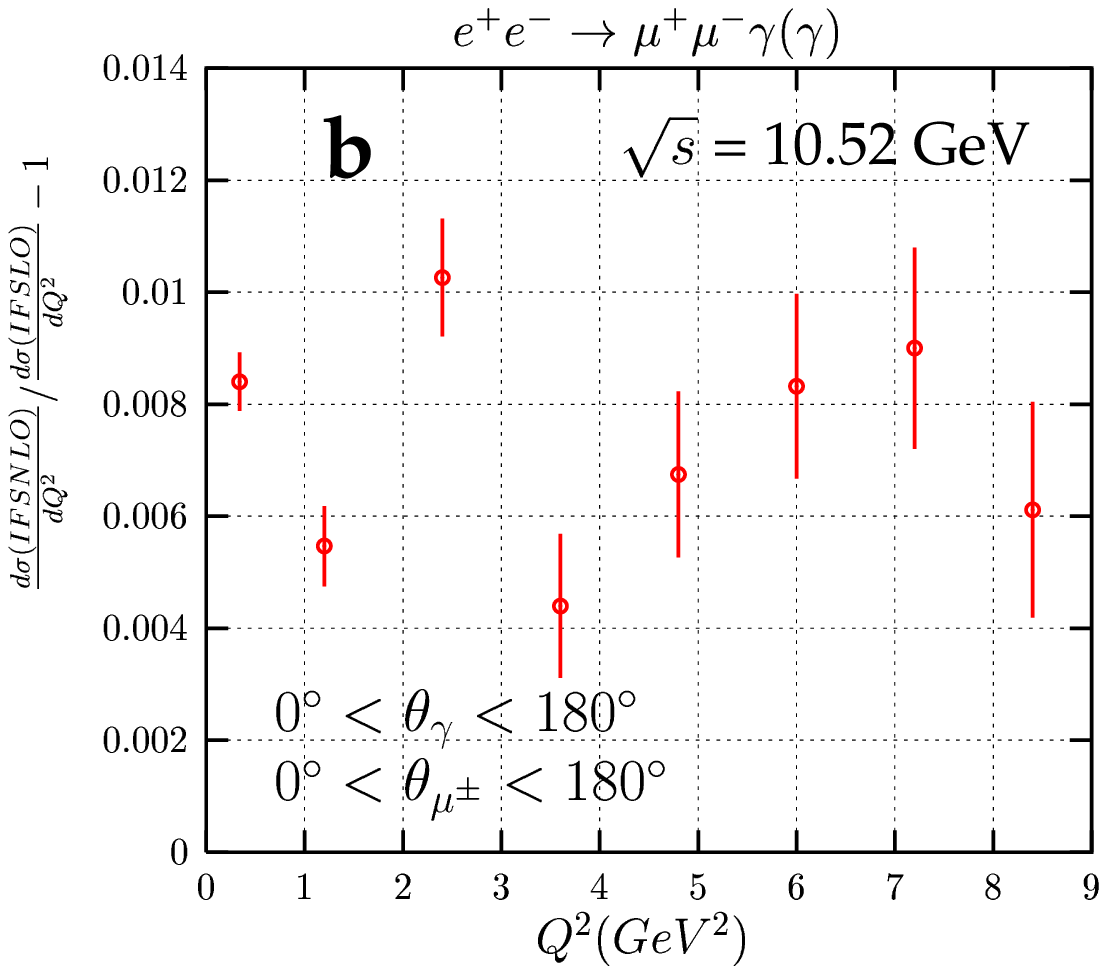,width=8.5cm} 
\epsfig{file=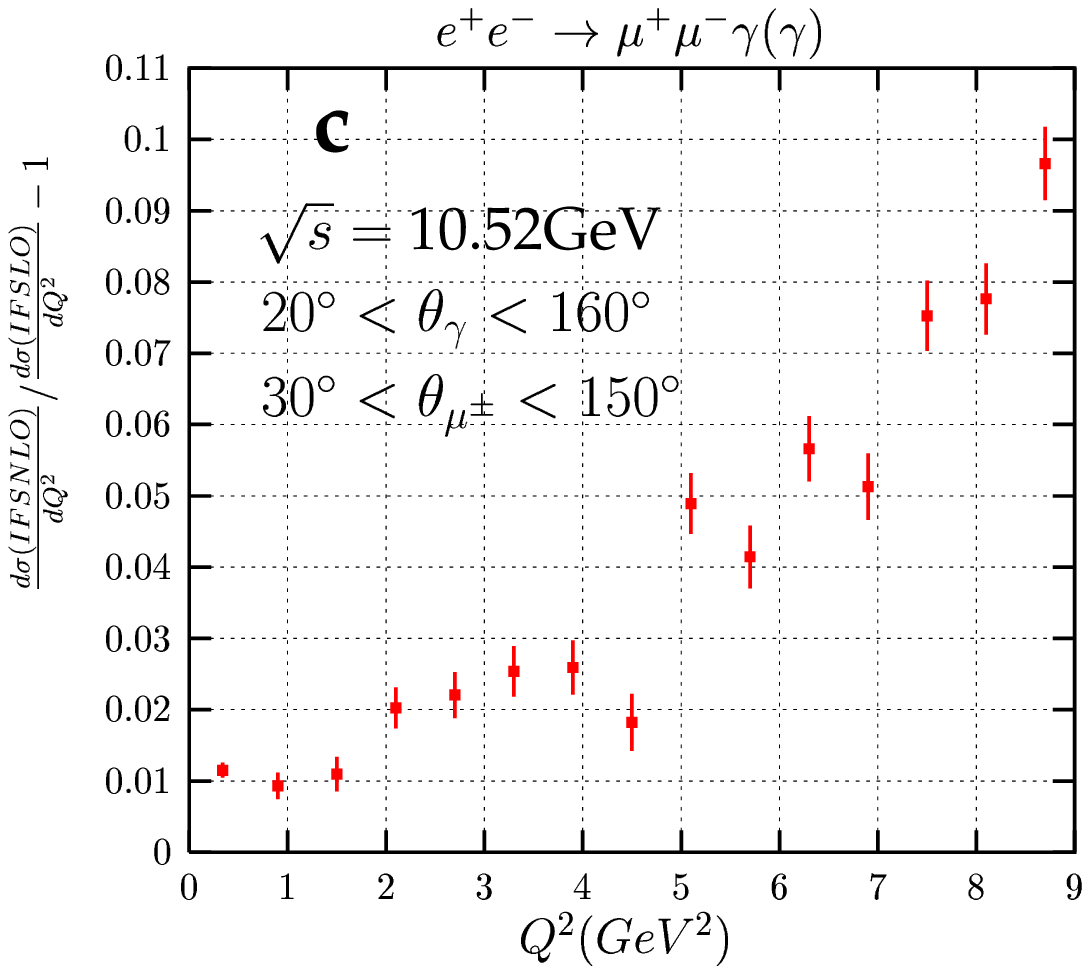,width=8.5cm} 
\caption{The size of the next-to-leading FSR corrections (IFSNLO) compared
with the sum of NLO ISR and LO FSR contributions (IFSLO) to the 
$e^+e^-\to \mu^+\mu^- \gamma(\gamma)$ cross section: 
a) $Q^2$ differential cross sections
with no angular cuts at $B$-factories;
b) relative difference of the cross sections with no angular cuts;
c) relative difference of the cross sections with angular cuts
resembling the BABAR detector (in $e^+e^-$ cms). 
If two photons are emitted, both hard (with energy $>$ 100~MeV) photons
are required to be within indicated angular cuts.}
\label{fig:10.52_sigma}
\end{center}
\end{figure}

At KLOE, the process $e^+e^-\to \mu^+\mu^-\gamma(\gamma)$ is a background 
to the  measurement of the  $e^+e^-\to \pi^+\pi^-$ cross section by the 
radiative return method, due to possible pion--muon misidentification. 
The angular cuts used by now by KLOE diminish the low $Q^2$ part of the 
cross section, while the track mass cut, which is specifically introduced 
to constrain the final state to configurations with $\pi^+\pi^-$ and one 
photon only (see Ref. \cite{Achim:radcor02}),
does not allow for events with $\mu^+\mu^-$ and one photon
only (see Fig.~\ref{fig:ang_trk}a for corresponding cross sections). 
If only angular cuts are applied, the newly introduced corrections
are at the level of a few per cent (Fig.~\ref{fig:ang_trk}b). However,
if in addition the track mass cut is applied, leaving only events with
two hard photons, the contribution of one-photon ISR plus one-photon FSR
to the two-photon sample is up to 35\% of the two-photon ISR sample 
(Fig.~\ref{fig:ang_trk}c), thus this contribution is indispensable
for a reliable background estimate.
The left-over two-hard-photon corrections, with both photons emitted from
the final state, are expected to be smaller, but will be relevant anyhow
if the precision requirement for the background estimate is of 
the order of a few per cent.

The process $e^+e^-\to \mu^+\mu^-\gamma(\gamma) $ is even
more important for BABAR, where it is used as $Q^2$-dependent 
luminosity monitoring. As a result the accuracy requirement is higher.
In Fig.~\ref{fig:10.52_sigma} some features of the newly implemented
corrections are shown in the interesting $Q^2$ range, from the point 
of view of using the radiative return method for measurements
of the hadronic cross section.
In Fig.~\ref{fig:10.52_sigma}a the $Q^2$ differential cross
section is given, including and excluding NLO
FSR corrections. As seen from  Fig.~\ref{fig:10.52_sigma}b the 
relative size of those corrections is quite small, of the order 
of 1\%, if no angular cuts are applied; but, as shown in 
Fig.~\ref{fig:10.52_sigma}c, their relevance depends on the event 
selection and the corrections can be as big as 10\% for some event 
selection and $Q^2$ range.


\section{ISR corrections to FSR real hard emission for two-pion final state}

\begin{figure}
\begin{center}
\epsfig{file=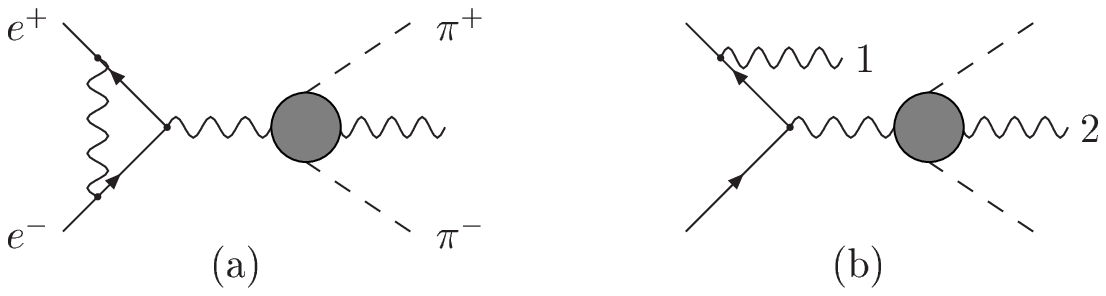,width=8.5cm} 
\caption{Virtual (a) and real (b) radiative corrections to real FSR
emission from two-pion final state.}
\label{pi_diagram}
\end{center}
\end{figure}

Initial-state vertex corrections to FSR for the reaction
$e^+e^- \to \pi^+\pi^-\gamma$ (Fig.~\ref{pi_diagram}a) were not included in 
version 3.0 of PHOKHARA, and therefore, for consistency, double photon 
radiation (Fig.~\ref{pi_diagram}b) was restricted to emission of hard ISR. 
In the present version of the event generator, PHOKHARA 4.0,
these corrections~\cite{Berends:1988np} have been added and the 
generation of double photon emission is unconstrained, provided that one 
of the photons is hard.
The virtual and soft corrections to the initial-state vertex are identical 
to the muon case and Eq.~(\ref{sigm_i}) holds, with 
$d\sigma^{(0)}_{\mathrm{FSR}}$ now being the leading order 
$e^+e^- \to \pi^+\pi^-\gamma$ cross section, and the photon 
being emitted off the final pions only.

\begin{figure}
\begin{center}
\epsfig{file=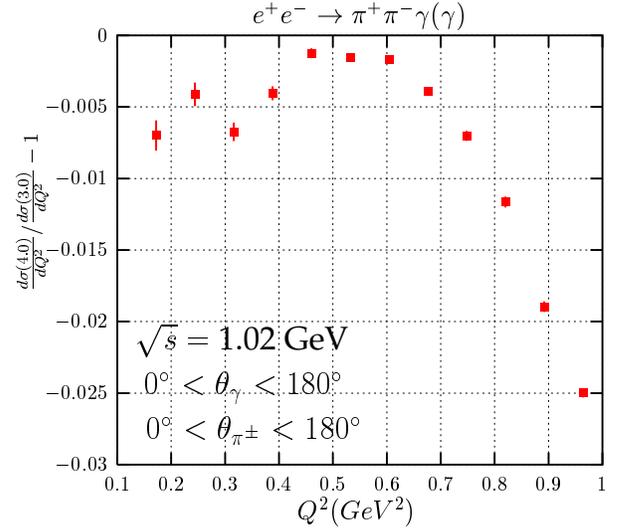,width=8.5cm} 
\caption{Comparison between PHOKHARA 3.0 and PHOKHARA 4.0
for $\sqrt{s} =$ 1.02~GeV. The pion and photon(s) angles 
are not restricted.}
\label{Comp_pi_all}
\end{center}
\end{figure}

We report here also formulae analogous to Eqs.~(\ref{etavs}) and 
(\ref{etavsp}), as we have found representations simpler than previously 
reported in~\cite{Czyz:PH03}. The virtual plus soft corrections, with 
soft photon energy cut $E_2^\mathrm{cut}$ in the $s'$ rest frame, read
\begin{align}
& \eta^\mathrm{V+S}(s',E_2^\mathrm{cut})  = 
- 2 \biggl[ \frac{1+\beta_{\pi}^2}
{2\beta_{\pi}} \log(t_{\pi}) + 1 \biggr] 
\log(2w)  \nonumber \\
&- \frac{2+\beta_{\pi}^2}{\beta_{\pi}} \log(t_{\pi}) - 2 
+ \log\biggl(\frac{1-\beta_{\pi}^2}{4}\biggr) 
\nonumber \\ &
- \frac{1+\beta_{\pi}^2}{2\beta_{\pi}} \biggl\{ -2\log(t_{\pi}) 
\log\biggl(\frac{1+\beta_{\pi}}{2}\biggr)
\nonumber \\ &
+ 4\Li_2 \biggl( \frac{2\beta_{\pi}}{1+\beta_{\pi}} \biggr)
- \pi^2 \biggr\}~, 
\label{etavs_pi}
\end{align}
where \ $w = E_{2}^{\mathrm{cut}}/\sqrt{s'}$, 
$\beta_{\pi} = \sqrt{1 - 4m_{\pi}^{2}/s'}$ 
and $t_{\pi} = (1-\beta_{\pi})/(1+\beta_{\pi})$,
while, with a soft photon energy cut 
$E_{\gamma}^\mathrm{min}$ in the $e^+e^-$ cms frame, they read
\begin{align}
& \eta^\mathrm{V+S}(s,s',E_{\gamma}^\mathrm{min})  = 
- 2 \biggl[ \frac{1+\beta_{\pi}^2}
{2\beta_{\pi}} \log(t_{\pi}) + 1 \biggr] 
\nonumber \\ & \times
\biggl[ \log(2w) + 1 + \frac{s'}{s'-s} 
\log\biggl(\frac{s}{s'} \biggr) \biggr] \nonumber \\
&- \frac{2+\beta_{\pi}^2}{\beta_{\pi}} \log(t_{\pi}) - 2 
+ \log\biggl(\frac{1-\beta_{\pi}^2}{4}\biggr) 
\nonumber \\ &
- \frac{1+\beta_{\pi}^2}{2\beta_{\pi}} \biggl\{ - \log(t_{\pi}) 
 \log\biggl(\frac{1+\beta_{\pi}}{2}\biggr) 
\nonumber \\ &
+ 4\Li_2 \biggl( \frac{2\beta_{\pi}}{1+\beta_{\pi}} \biggr)
- \pi^2 \biggr\}~, 
\label{etavsp_pi}
\end{align}
where $w = E_{\gamma}^\mathrm{min}/\sqrt{s}$. 

\begin{figure}
\begin{center}
\epsfig{file=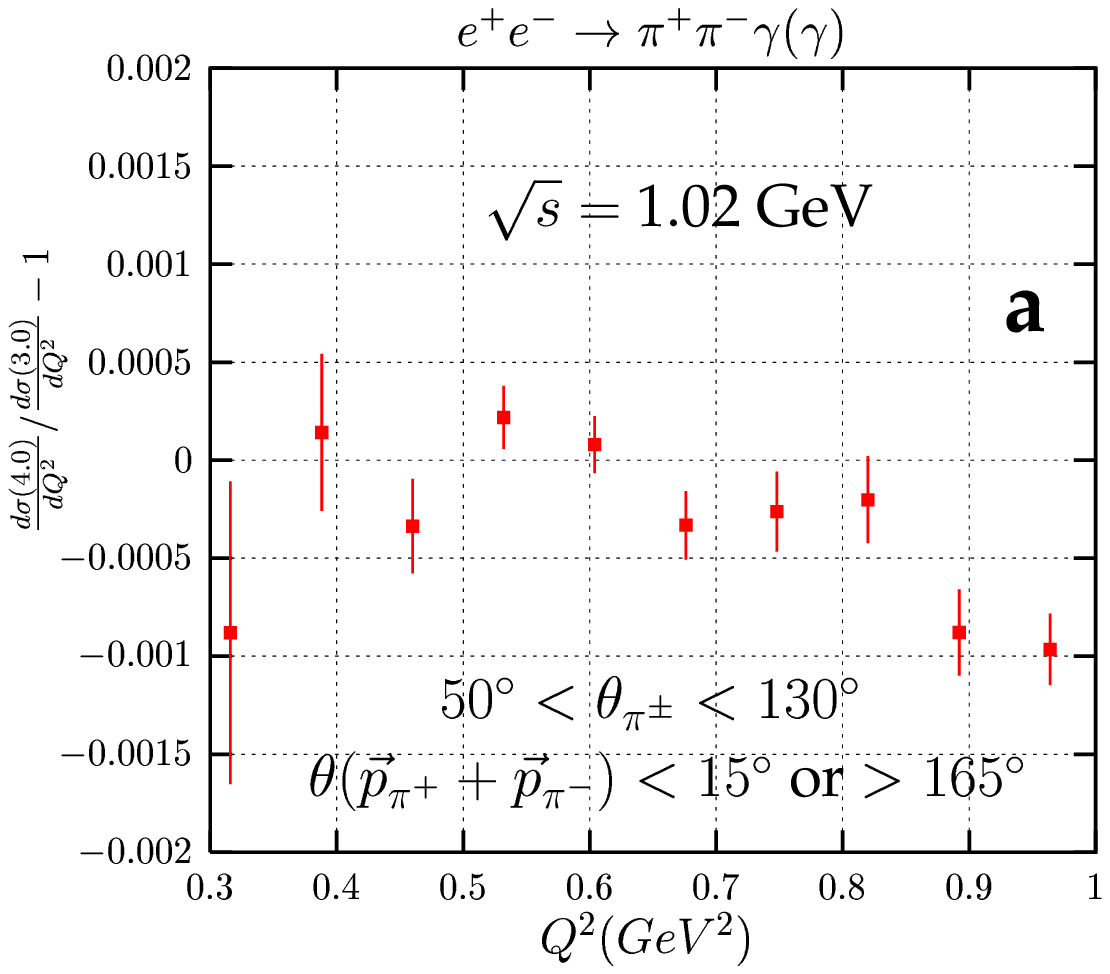,width=8.5cm} 
\epsfig{file=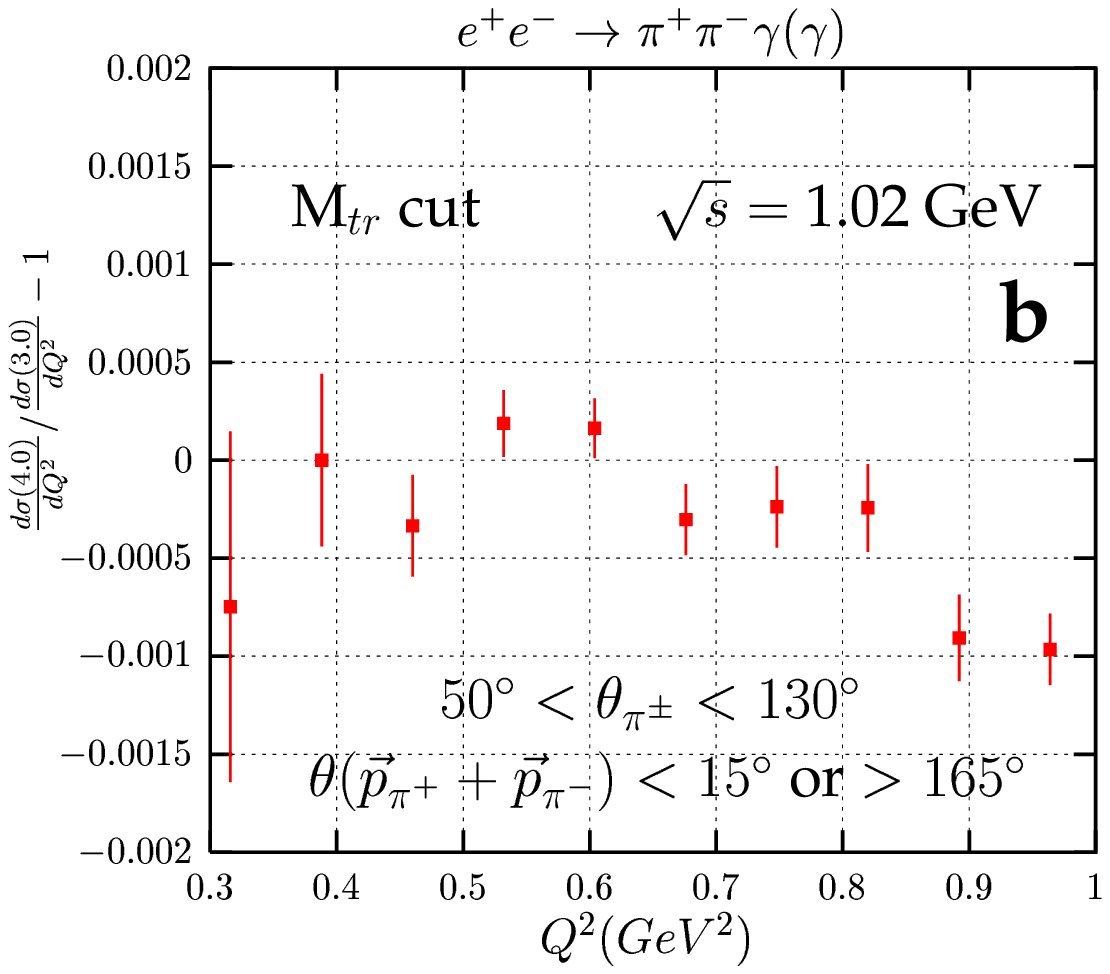,width=8.5cm} 
\caption{Comparison between PHOKHARA 3.0 and PHOKHARA 4.0 for 
$\sqrt{s}=1.02$~GeV: a) the pion polar angles are restricted to 
50$^\circ < \theta_{\pi^\pm} < $ 130$^\circ $ 
and angular cuts are imposed on the missing 
momentum; b) the same as (a) with additional track mass cut.}
\label{Comp_pi_ang}
\end{center}
\end{figure}

\begin{figure} 
\begin{center}
\epsfig{file=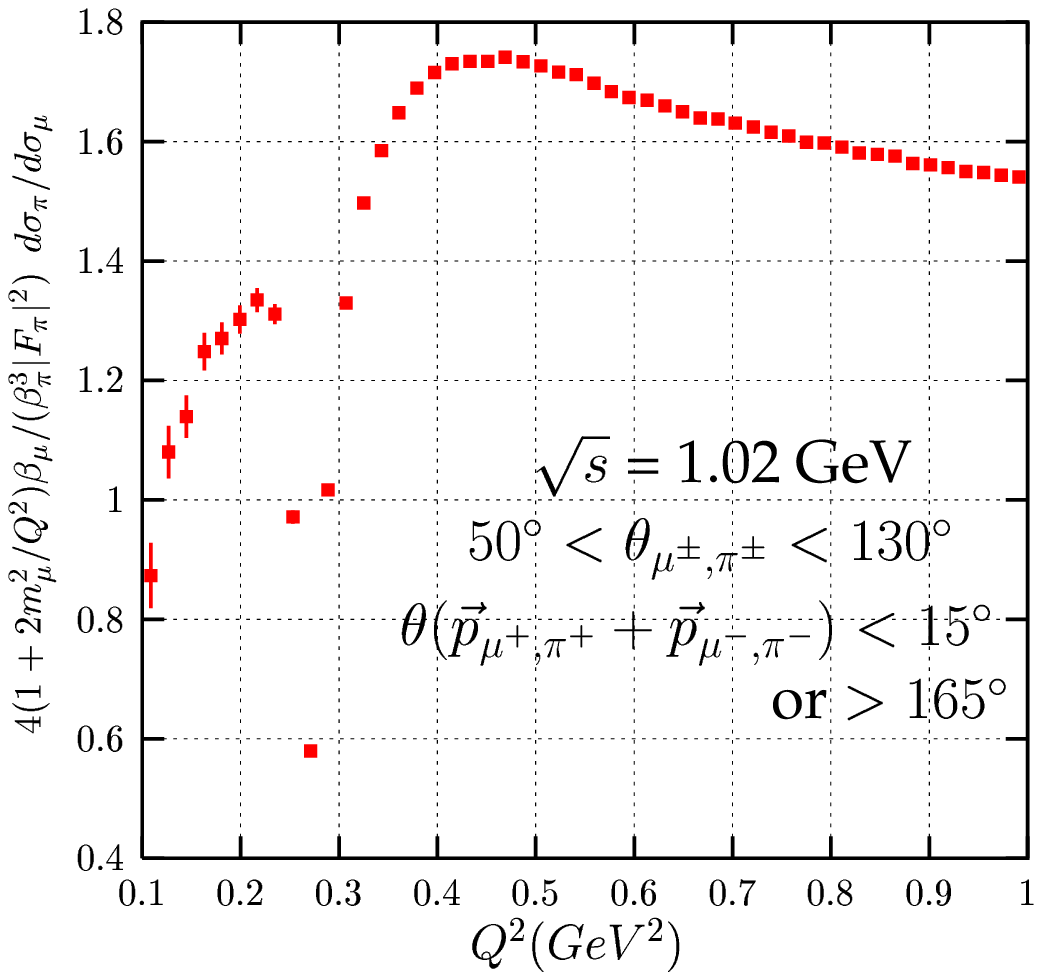,width=8cm} 
\epsfig{file=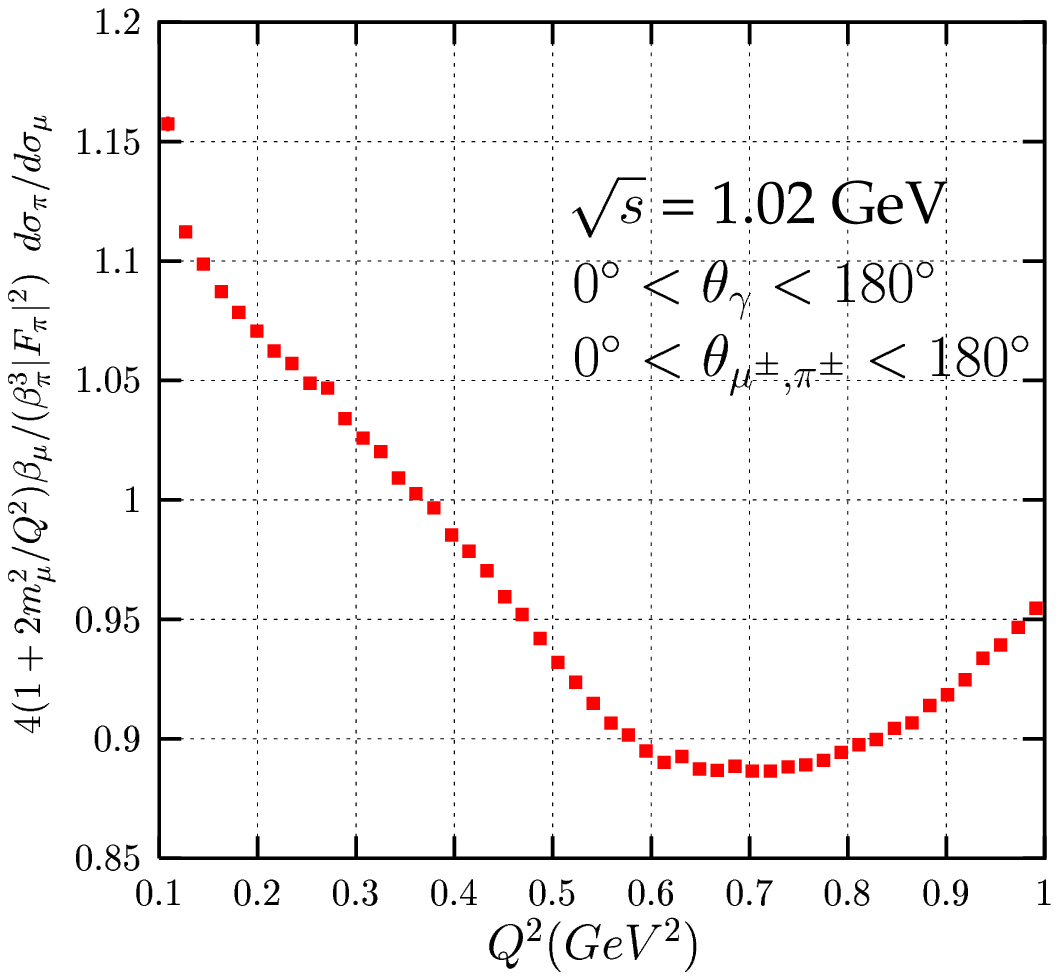,width=8cm} 
\epsfig{file=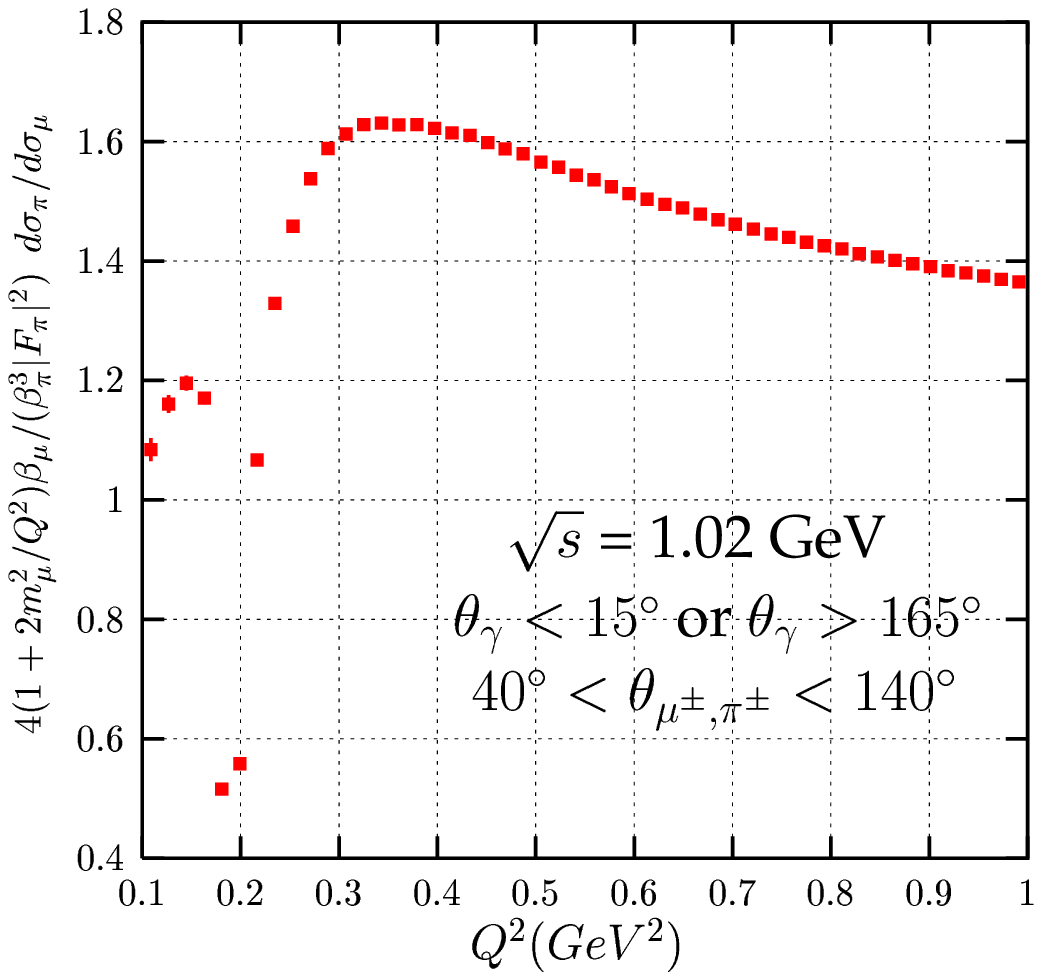,width=8cm} 
\caption{The ratio $\cal R$ of Eq.~(\ref{rmupi}) for $\sqrt{s}$ = 1.02~GeV,
for three different event selections.}
\label{fig:mu_pi_50}
\end{center}
\end{figure}

\begin{figure} 
\begin{center}
\epsfig{file=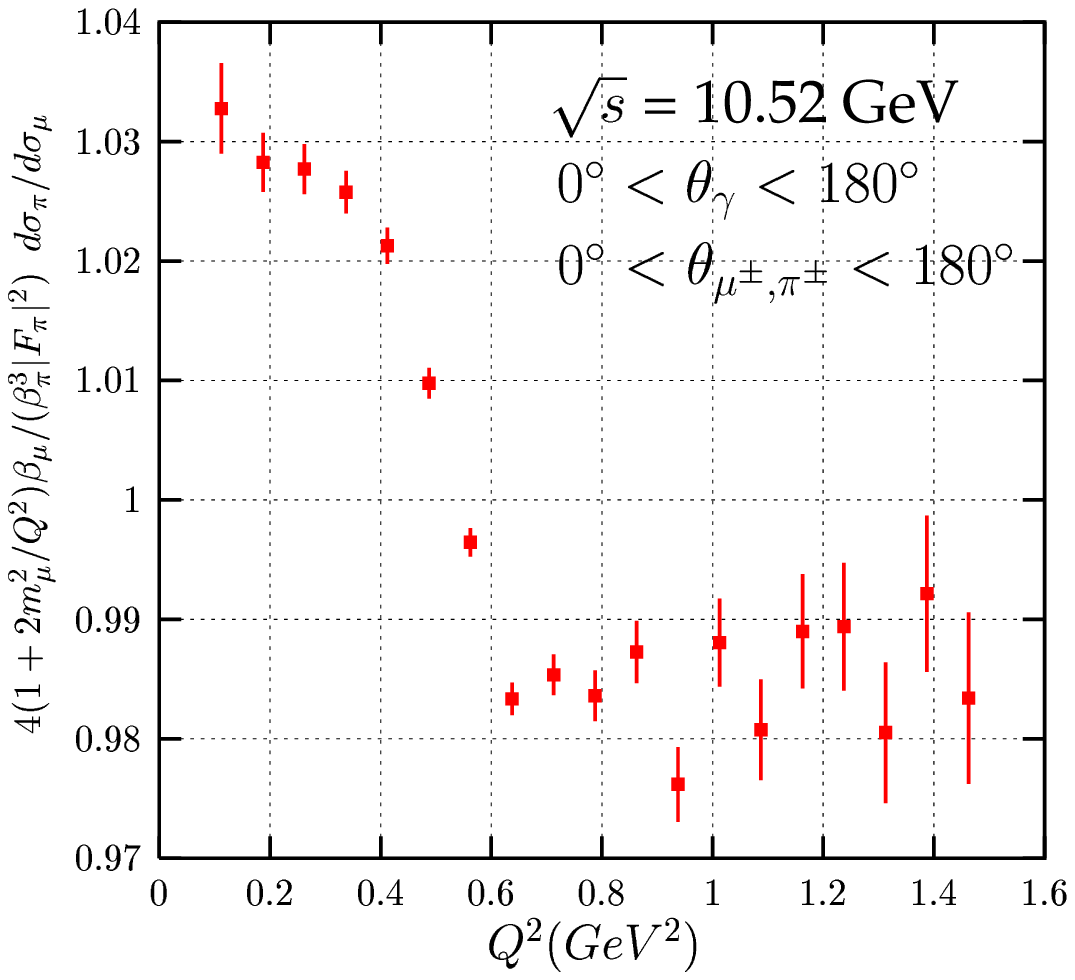,width=8cm} 
\epsfig{file=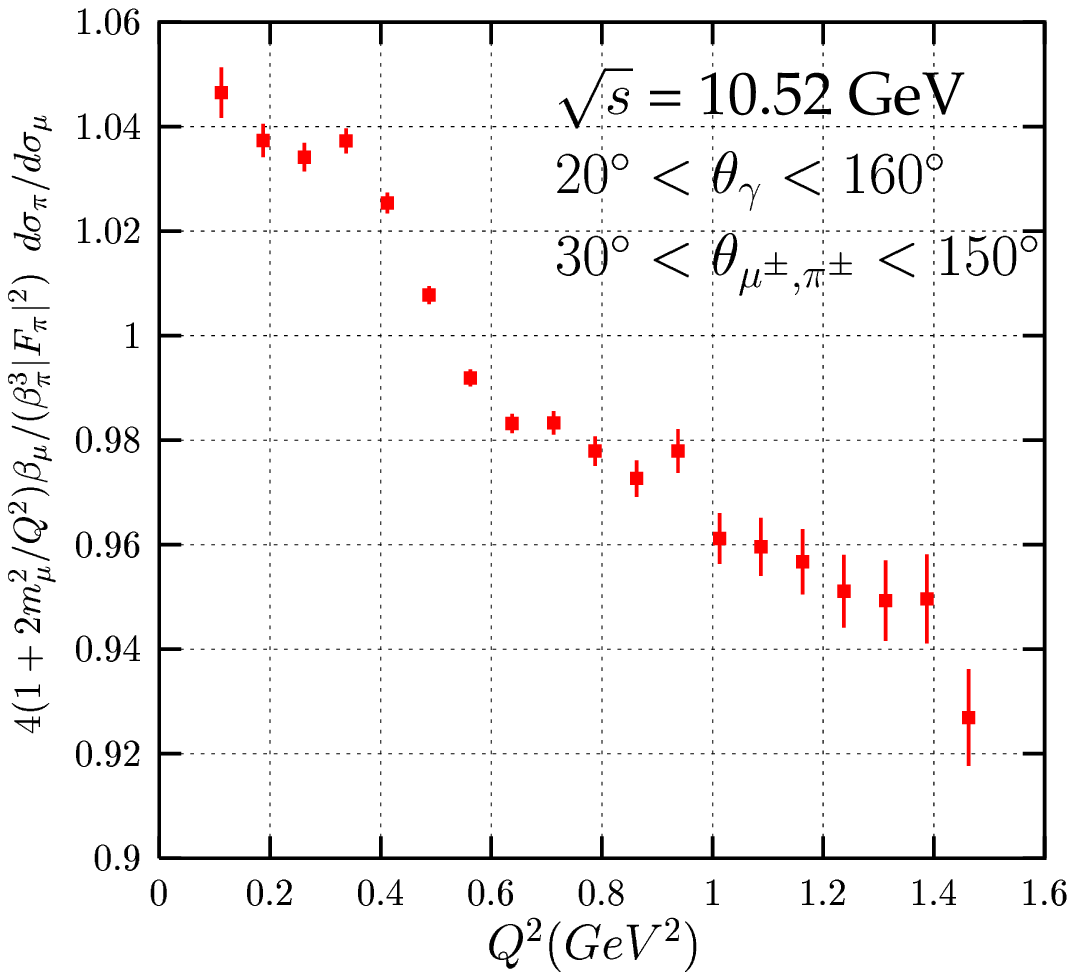,width=8cm} 
\caption{The ratio $\cal R$ of Eq.~(\ref{rmupi}) for $\sqrt{s}$ = 10.52~GeV
and two different event selections.}
\label{fig:10.52_out_mupi}
\end{center}
\end{figure}

The numerical importance of the corrections from   
Fig.~\ref{pi_diagram} depends on the event selection.
If no angular cuts are imposed they contribute up to 2--3\% 
to the $Q^2$ differential cross section, as shown in Fig.~\ref{Comp_pi_all}.
However, for KLOE cuts used for low-angle measurements of the pion form factor,
the additional IFSNLO corrections, not included in PHOKHARA 3.0, are well
below 0.1\%, as anticipated in~\cite{Czyz:PH03}. It is shown 
in Fig.~\ref{Comp_pi_ang} for event selections with and without the track mass
cut. It is worth mentioning that the IFSNLO corrections change also the ratio
\bea
{\cal R}(Q^2)
 = \frac{4(1+\frac{2m_\mu^2}{Q^2})\beta_\mu}{\beta_\pi^3|F_\pi(Q^2)|^2}
 \frac{\frac{d\sigma_\pi}{dQ^2}}{\frac{d\sigma_\mu}{dQ^2}} \ ,
\label{rmupi}
\eea
which is expected to be 1 for ISR only, when no cuts are applied. From 
Figs.~\ref{fig:mu_pi_50} and~\ref{fig:10.52_out_mupi} it is clearly
visible that one has to take the IFSNLO corrections into account and use a Monte
Carlo event generator, when extracting the pion form factor from the ratio
of pion and muon cross sections. This is true at both $\Phi$- and $B$-factories.
A similar situation can be expected for other hadronic final states.

\section{Vacuum polarization implementation}

\begin{figure}
\begin{center}
\epsfig{file=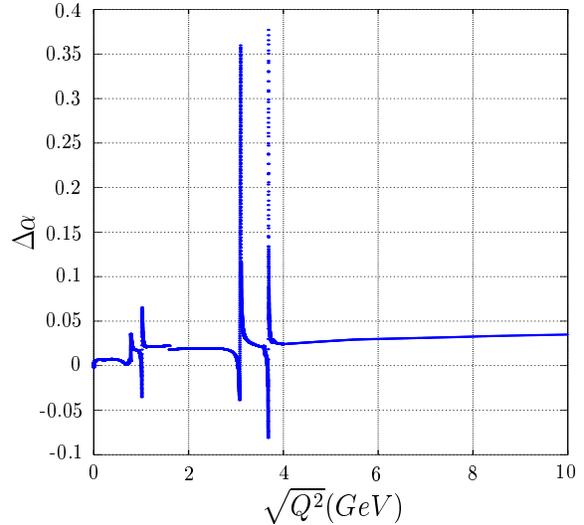,width=8cm} 
\caption{The energy dependence of $\Delta \alpha$ calculated by the program
of Ref.~\cite{Jeg_web}. }
\label{vac_1}
\end{center}
\end{figure}

\begin{figure}
\begin{center}
\epsfig{file=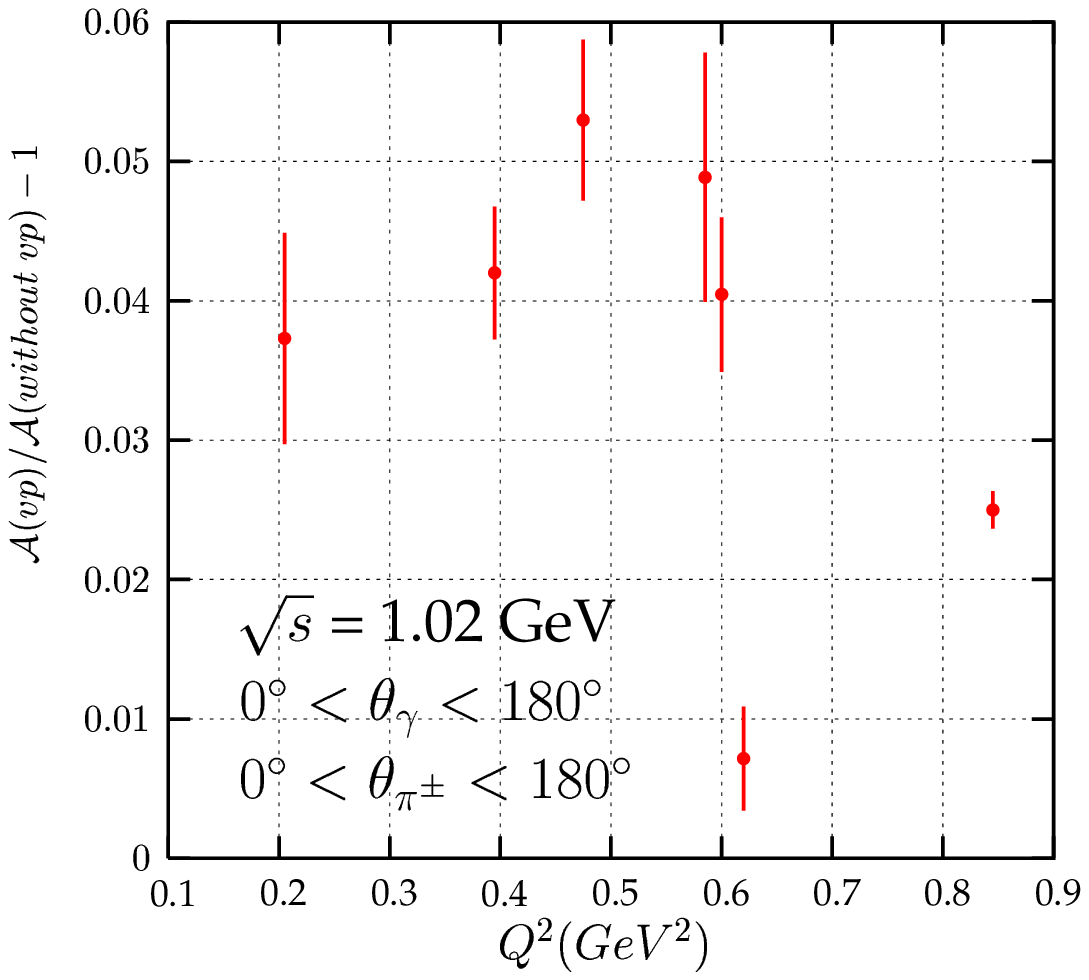,width=8.5cm} 
\caption{The relative difference of the forward--backward asymmetry
of the pion polar angle distribution, calculated with and without
vacuum polarization contributions.}
\label{vac_2}
\end{center}
\end{figure}

PHOKHARA 4.0 includes also vacuum polarization corrections for all
final hadronic states and muon production, with the actual implementation
of Ref.~\cite{Jeg_web}. As can be anticipated from Fig.~\ref{mu_diagram},
the vacuum polarization corrections are calculated at different
scales for different types of diagrams ($\hat Q^{2}$, $Q^{2}$, $s$, etc.).
In the energy range that is important for measurements of the components
of the hadronic cross section through the radiative return method, the vacuum 
polarization corrections have a non-trivial behaviour 
(see Fig.~\ref{vac_1}) due to the hadronic contributions. For event
selections, for which only ISR corrections are important, the
vacuum polarization contribution is just a multiplicative 
factor depending on $Q^2$, so that one can correct the event generator 
results even after generation.
However, for event selections where other contributions are also important,
event generation with complete implementation of vacuum polarization might
become necessary. An example is shown in Fig.~\ref{vac_2}, where the 
vacuum polarization contributions do change the forward--backward asymmetry
of the pion polar angle distribution. The effect, of up to 5\%, is mainly
caused by the choice of $s$ in the close vicinity of the $\Phi$ resonance,
which is nothing but the KLOE case.
As seen in Fig.~\ref{vac_1} the vacuum polarization around the $\Phi$ is much
larger than at lower energies. As a result, the vacuum polarization
correction to the ISR--FSR interference, which contributes to the numerator
of the asymmetry, is larger than the corresponding corrections 
to its denominator, which is dominated by ISR.
Therefore, vacuum polarization corrections do not drop in the ratio.
This effect has to be taken into account, when studies of model
dependence of FSR are performed.

\section{Summary and Conclusions}

Precision measurements of the hadronic cross section through the radiative
return to an accuracy of about 1\% require the full understanding of
radiative corrections and their implementation in a Monte~Carlo event generator.
For this reason a number of new features have been introduced 
into the Monte~Carlo event generator PHOKHARA.

In a first step we presented the formulae for real and virtual radiation, 
which are required to describe FSR for
muon pair production through the radiative return in next-to-leading order.
This includes a specific set of virtual corrections as well as real
radiation. The Monte~Carlo event generator PHOKHARA has been upgraded and, in its
present form, also simulates the contribution from the ``two-step'' reaction
$e^+e^-\to \gamma \gamma^* (\to \mu^+\mu^-)$. The technical precision of the
generator in this new mode has been demonstrated to be better than $10^{-4}$.
It is shown that the relative importance of this NLO correction depends
strongly on the details of the experimental cuts.

The program has also been extended to the case of pion production. 
In particular corrections from virtual plus soft ISR, 
in combination with hard FSR, have been
introduced. Although the impact of this additional piece is small, as far as
actual measurements at KLOE or the $B$-factories are concerned, it is required
for a complete NLO treatment of the radiative return for pion-pair production.
Another straightforward upgrade is the introduction of vacuum polarization in
the photon propagator. As far as charge-symmetric cross section measurements
are concerned, this effect can just be considered as a multiplicative,
$Q^2$-dependent change of the normalization. However, interference terms
between amplitudes with virtual photons of different virtualities, which are
relevant to e.g. the forward--backward asymmetry, do lead to modifications of
the results at the level of several per cent.


\section*{Acknowledgements}

We would like to thank A.~Denig, W.~Kluge, D.~Leone and S.~M\"uller for 
discussions of the experimental aspects of our analysis. 
H.C. and A.G. are grateful for the support and the kind hospitality of
the Institut f{\"u}r Theoretische Teilchenphysik of the Universit\"at Karlsruhe.
Special thanks to Suzy Vascotto for careful proof-reading of the ma\-nu\-script.


\end{document}